\documentclass{article}
\usepackage[utf8]{inputenc}
\usepackage{graphicx}
\usepackage{amssymb}
\usepackage{amsmath}
\usepackage{cmll}
\usepackage{booktabs}
\usepackage{ dsfont }
\usepackage{caption}
\usepackage{cite}  

\usepackage{dcolumn}
\usepackage{bm}

\usepackage{mathptmx}
\usepackage{etoolbox}
\usepackage{booktabs}

\usepackage[utf8]{inputenc} 
\usepackage{geometry}
\usepackage[T1]{fontenc}    
\usepackage{hyperref}       
\usepackage{url}            
\usepackage{booktabs}       
\usepackage{amsfonts}       
\usepackage{nicefrac}       
\usepackage{microtype}      
\usepackage{lipsum}		
\usepackage{doi}
\usepackage{indentfirst}
\usepackage{multirow}
\usepackage{float}
\usepackage{subfigure}
\usepackage{graphicx}
\usepackage{float}

\renewcommand{\thesubfigure}{(\roman{subfigure})}
\makeatletter \renewcommand{\@thesubfigure}{\thesubfigure \space}
\renewcommand{\p@subfigure}{} \makeatother

\title{Model Based Reinforcement Learning with Non-Gaussian Environment Dynamics and its Application to Portfolio Optimization}

\author{\bf\normalsize{
Huifang Huang$^{1,}$,
Ting Gao$^{1,}$\footnote{\url{tgao0716@hust.edu.cn}},
Pengbo Li$^{1,}$,
Jin Guo$^{1,}$,
Peng Zhang$^{1,}$,
Nan Du$^{2,}$
}\\[10pt]
\footnotesize{$^1$ School of Mathematics and Statistics \& Center for Mathematical Sciences.} \\
\footnotesize{Huazhong University of Science and Technology, Wuhan 430074, China.} \\[5pt]
\footnotesize{$^2$ Tencent AI Lab, Shenzhen, China.}
}

\begin{document}
\maketitle

\begin{abstract}

\noindent With the fast development of quantitative portfolio optimization in financial engineering, lots of AI-based algorithmic trading strategies have demonstrated promising results, among which reinforcement learning begins to manifest competitive advantages. However, the environment from real financial markets is complex and hard to be fully simulated, considering factors such as abrupt transitions, unpredictable hidden causal factors, heavy tail properties and so on. Thus, in this paper, first, we adopt a heavy-tailed preserving normalizing flows to simulate high-dimensional joint probability of the complex trading environment and develop a model-based reinforcement learning framework to better understand the intrinsic mechanisms of quantitative online trading. Second, we experiment with various stocks from three different financial markets (Dow, NASDAQ and S\&P) and show that among these three financial markets, Dow gets the best performance based on various evaluation metrics under our back-testing system. Especially, our proposed method is able to mitigate the impact of unpredictable financial market crises during the COVID-19 pandemic, resulting in a lower maximum drawdown. Third, we also explore the explanation of our RL algorithm. (1), we utilize the pattern causality method to study the interactive relation among different stocks in the environment. (2), We analyze the dynamic loss and actor loss to ensure the convergence of our strategies. (3), by visualizing high dimensional state transition data comparisons from real and virtual buffers with t-SNE, we uncover some effective patterns of better portfolio optimization strategies. (4), we also utilize eigenvalue analysis to study the convergence properties of the environment's model.
\end{abstract}

\section{Introduction}
With the development of modern deep learning techniques, more and more cutting-edge neural network structures have been widely used in various research and application fields. These machine learning frameworks in return inspire researchers from different backgrounds to design better neural network architectures to improve the model performance and solve more complex scenarios. One such promising application-quantitative finance-has recently been taking advantage of AI techniques to make lots of innovative algorithmic trading strategies for financial investment and portfolio optimization. 


 Among the machine learning frameworks, reinforcement learning (RL) has a unique advantage in that the interactive training process is aligned with human decision-making. Over the past years, two basic optimization-Q-learning and Policy gradient-have been developed to solve many RL problems, as well as the combination of the two. Take financial applications for example: Pastore et al.~\cite{2016Modelling} analyzed data on 46 players from online games in financial markets and test whether the Q-Learning could capture these players' behaviors based on a riskiness measure. Their results indicated that not all players were short-sighted, which contradicts the naive-investor hypothesis. Additionally, a paper done by, Li et al. (2019)~\cite{2019An} studied the benefits of the three different classical deep RL models (DQN~\cite{mnih2013playing}, Double DQN~\cite{van2016deep} and Dueling DQN~\cite{wang2016dueling}) by predicting the stock price and concluded that DQN had the best performance. In addition, some researchers simulated and compared the improved deep RL method with the Adaboost algorithm and proposed a hybrid solution. Interestingly, Lee et al. (2019)~\cite{2019Global} applied CNN to a deep Q-network that takes stock price and trading volume images as inputs to predict global stock markets. On the policy gradient side, Kang et al. (2018)~\cite{2018An} utilized the state-of-art Asynchronous Advantage Actor-Critic algorithm (A3C~\cite{mnih2016asynchronous}) to solve the portfolio management problem and designed a standalone deep RL model. Moreover, Li et al. (2019)~\cite{2019Optimistic} propose an adaptive deep deterministic RL method (Adaptive DDPG) for some portfolio allocation tasks, which incorporated optimistic or pessimistic deep RL algorithm based on the influence of prediction error. Through analyzing data from Dow 30 component stocks, they showed that the trading strategy outperforms the traditional DDPG method~\cite{lillicrap2015continuous}. Sutta (2019)~\cite{sornmayura2019robust} compared the performance of an AI agent to the results of the buy-and-hold strategy and the expert trader by testing 15 years' forex data market with a paired t-Test. The findings showed that AI could beat the buy \& hold strategy and commodity trading advisor in FOREX for both EURUSD and USDJPY currency pairs. However, these algorithms are quite expensive to train, considering the high risk of investment loss at an early stage in the real financial market environment, even using simulated domains (Mnih et al. (2015)~\cite{mnih2015human}, Schulman et al. (2017)~\cite{schulman2017proximal}). A promising direction for improving sample efficiency is to explore model-based reinforcement learning (MBRL) methods~\cite{yu2019model}. Chua, Calandra et al. (2019)~\cite{chua2018deep} studied how to bridge this gap by employing uncertainty-aware dynamical models. They proposed a novel algorithm called PETS, which approaches the asymptotic performance of several benchmark model-free algorithms while requiring significantly fewer samples. Janner et al. (2019)~\cite{janner2019trust} proposed model-based Policy Optimization (MBPO) algorithm to boost PETS through monotonic improvement at each step to get the best performance.

However, reinforcement learning is still plagued with intricate and difficult issues in the financial world. Firstly, unforeseen variables such as international conditions, national policies, and business news events have an impact on financial markets. Secondly, hidden interactions among different agencies can be unpredictably difficult to understand. As shown from lots of stock marketing analyses, agencies with large capital investments can sometimes cause dramatic price fluctuation, leading to marketing panic and instability among the public. Additionally, complex nonlinearity, chaotic behaviors, as well as non-Gaussian noise tend to be present in financial marketing data, which results in a distribution shift~\cite{cai2020distributed} of financial time series data over time. Complex causal relationships exist among different stocks. Stavroglou (2019)~\cite{nf_stavroglou2019hidden} characterized the interaction of the local spatiotemporal dynamics between the attractors of asset pairs ($M_X$ and $M_Y$) as pattern causality (PC). In this study, we empirically examine using the PC technique if there is a meaningful causal relationship between experimental stocks from various marketplaces and whether the magnitude of that relationship affects them. Many efforts have been made by academics to address the processes behind the aforementioned issues. On one hand, various data denoising methods have been utilized to preprocess the financial time series data. For example, Bao W, et al. (2017)~\cite{2017A} proposed a new model for stock prediction, wavelet transforms for decomposing and eliminating the noise of the stock price time series, which helps stacked auto-encoders (SAEs) for extracted high-level deep features and long-short term memory (LSTM) to better forecast the stock price and improve predictive accuracy and profitability performance. On the other hand, researchers have also created many kinds of indexes to optimize portfolio gain. In short, all of these considerations make it hard to build a reinforcement learning system that directly interacts with a complex real marketing environment, which motivates us to investigate model-based RL for algorithmic trading strategy optimization.

Despite the fact that stock prices are noisy and nonlinear, the difference between two consecutive days' stock prices could be considered as a stationary distribution (Ariyo et al. (2014)~\cite{ariyo2014stock}). In this article, the stock price manifests as non-Gaussian $\alpha$-stable l$\acute{e}$vy noise. Therefore, in addition to the Classical MBRL algorithms, such as PETS and MBPO, which assume that each state component is independent of the others and follows a Gaussian distribution, we need to explore an environment that better depicts the non-Gaussian properties. This inspires us to seek some heavy-tailed presenting normalizing flows.

By building a series of bijective transformations, normalizing flows (Dinh (2014)~\cite{dinh2014nice}; Dinh (2016)~\cite{dinh2016density}) proposed a mechanism for producing high-dimensional data distributions. Significant promise has already been shown in the creation and identification of images. Papamakarios et al.(2021)~\cite{papamakarios2021normalizing} described flows using probabilistic modeling and inference. Kashif Rasul et al.(2020)~\cite{rasul2020multivariate} offered autoregressive deep learning architectures for simulating the temporal dynamics of multivariate time series using normalizing flows. Mevlana et al.(2016)~\cite{gemici2016normalizing } considered the problem of density estimation on Riemannian manifolds by employing normalizing flows.

In this paper, We first simulate one high-dimensional stock state space with an $\alpha$-stable l$\Acute{e}$vy distribution and propose a strategy model-based normalizing flows (MBNF) by employing normalizing flows as the dynamic model. We also compare the performance of MBNF with MBPO method in three markets and give the algorithm explanation by utilizing the pattern causality method, eigenvalue analysis and the t-SNE method.

Overall, this paper makes the following major contributions: 
\begin{itemize}
    \item\textbf{Heavy-tailed Distribution as Transition States}. 
    We adopt normalizing flows to simulate the high-dimensional joint probability of the complex trading environment, which can capture the interactive relationships among some stocks from upstream and downstream enterprises.
   
    \item\textbf{Business Performance}.
    We experiment with various stocks from three different financial markets (Dow, NASDAQ and S\&P) with MBNF, and we show all these results of MBNF are comparatively better than modeling the state transition dynamics with independent Gaussian Processes in terms of seven back-testing metrics.

    \item\textbf{Explanable Neural Nework}.
    We utilize the pattern causality method to study the causal relationship among different stocks of the environment. Furthermore, we study the convergence landscape of the transition dynamic model through eigenvalue analysis. We also uncover some effective buffer data patterns of better portfolio optimization strategies. 
\end{itemize} 

We structure our paper in the following sessions: First, in Section 2.1, we introduce some background knowledge of model-based RL and then explain our financial environment setup under the model-based RL framework in Sections 2.2 and 2.3. We then interpret the transition dynamic from a dynamical system point of view and describe the Gaussian Process as some stochastic differential equation~(SDE), and correspondingly, the policy optimization of SAC~\cite{haarnoja2018soft} is connected with stochastic optimal control under the constraint of an SDE in Section 2.4. Next, we explain the normalizing flows method in detail in Section 3 and present the pseudo-code of our proposed model. All the experiment results are shown in Section 4, as well as detailed explanations of how and why our proposed algorithm is better than classic MBPO. Finally, we summarize our conclusions and future research directions in Section 5.

\section{Model Framework}
Reinforcement learning aims to learn an optimal policy so as to maximize the expectation of cumulative rewards in the process of interacting with the environment.


\subsection{Definition}
When considering a financial market in an RL environment, it is natural to treat a trading platform as the intelligent agent, with whom the stock trading can be modeled as a Markov Decision Process (MDP). Note that in this paper, we use bold to represent vectors.

$\bullet \textbf{State space}$ $\mathbb{S} = [B, \mathbf{P}, \mathbf{W}, \mathbf{I}]$:
State space $\mathbb{S}$ is the collected market information. $\forall s_t \in \mathbb{S}$ is the state of the agent at time $t$, which includes four parts of information: account balance for agency $B_t$, current stock price $\mathbf{P_t}$, cumulative holding $\mathbf{W_t}$, and technical indicators $\mathbf{I_t}$. Here, the technical indicators $I_t$ is consisted of seven common technical indicators: MACD~\cite{appel2003become}, SMA30, SMA60, BOLL~\cite{bollinger1992using}, RSI~\cite{ctuaran2011relative}, CCI~\cite{lai2020bidirectional}, and ADX. Setting d stocks in each market, Table \ref{table_1} describes the notations of state $s_t$ in detail.

\begin{table}
\caption{Notations of State Space}
\label{table_1}
\centering
\begin{tabular}{l|lll} 
\hline
Notation & Definition \\ 
\hline
$B_t$ & \begin{tabular}[c]{@{}l@{}} Account cash at time t; $B_t \in \mathbb{R}_+$ \end{tabular} \\
\hline
$\mathbf{P_t}$ & \begin{tabular}[c]{@{}l@{}} daily closing price at time t; $\mathbf{P_t} \in \mathbb{R}_+^{d}$ \end{tabular} \\
\hline
$\mathbf{W_t}$ & \begin{tabular}[c]{@{}l@{}} cumulative holding shares of each stock; $\mathbf{W_t} \in \mathbb{Z}_+^{d}$ \end{tabular} \\
\hline
MACD & \begin{tabular}[c]{@{}l@{}} Moving Average Convergence Divergence: \\a momentum indicator displays trend \end{tabular} \\
\hline
SMA30 & \begin{tabular}[c]{@{}l@{}} 30 day Simple Moving Average:\\ 30 day closing price equal-weighted average \end{tabular} \\
\hline
SMA60 & \begin{tabular}[c]{@{}l@{}} 60 day Simple Moving Average: \\60 day closing price equal-weighted average \end{tabular} \\
\hline
BOLL & \begin{tabular}[c]{@{}l@{}} Bollinger Bands:\\ judges the medium and long-term movement trend \end{tabular} \\
\hline
RSI & \begin{tabular}[c]{@{}l@{}} Relative Strength Index: \\identifies inflection points of a trend \end{tabular} \\
\hline
CCI & \begin{tabular}[c]{@{}l@{}} Commodity channel index: \\helps to find the degree of deviation of the price \end{tabular} \\ 
\hline
ADX & \begin{tabular}[c]{@{}l@{}} Average Directional Index:\\ determines the strength of a trend \end{tabular} \\
\hline
\end{tabular}
\end{table}

$\bullet$ \textbf{Action Space} $\mathbb{A}$: Action space $\mathbb{A}$ is a set of available operations during the transaction on d stocks. The action taken by the agent at time $t$ denoted as $\mathbf{a_t}$, is assumed to be finite-dimensional and continuous. Here, $\mathbf{a_t} \in \mathbb{A}$ is a d-dimensional vector, where the $i_{th}$ dimension represents the action performed on the $i_{th}$ stock.
 
Denoting $\mathbf{W^i_t}$ as the cumulative shares of $i_{th}$ stock at time $t$, we assume the maximum trading volume for a single stock at one step is 100 shares. The available actions of the stock include buying, holding and selling:
• Buying: $a^i_t = + h$, $h$ shares can be bought which leads to $\mathbf{W}^i_{t+1} = \mathbf{W}^i_t + h$, where $h \in [0,100]$ is a positive integer.\\
• Selling: $a^i_t = - h$, $h$ shares can be sold from the current holdings. In this case, $\mathbf{W}^i_{t+1} = \max( \mathbf{W}^i_t - h, 0)$.\\
• Holding: $a_t = 0$, which means no change in $\mathbf{W}^i_t$.

$\bullet$ \textbf{Reward} $r_t$: The reward at time $t$ is a mapping $ r_t: \mathbb{S} \times \mathbb{A} \rightarrow \mathbb{R}$, denote $r_t = r (s_t, a_t)$. In this work, we use the daily change amount of the net account value as $r (s_t, a_t)$, which is an intuitive definition for financial markets. The specific formula is shown in equation (3):

1)The total amount of the agent's net account value at time $t$, denoted by $Asset_t$, is the sum of current cash and the value of stock holdings. That is:
\begin{equation}
    Asset_t = B_t + \mathbf{P_t}^\prime \cdot \mathbf{W_t}
         = B_0 - \sum_{\tau = 0}^t \mathbf{P_{\tau}}^\prime \cdot a_{\tau} + \mathbf{P_{t+1}}^\prime \sum_{\tau = 0}^t a_{\tau}
\end{equation}
Where $\mathbf{P_t}^\prime$ is the vector transposition.

2) Considering the transaction cost at time $t$, we denote the cost as $C_t$:
\begin{equation}
    C_t = \mathbf{P_t}^\prime \cdot|\mathbf{a_t}|\cdot cost_{percentage}
\end{equation}
Where $|\mathbf{a_t}|$ means the absolute value of each component of the vector $\mathbf{a_t}$.

3) We define the reward $r_t$ in the following form:
\begin{equation}
    r_t = r(s_t, a_t)= Asset_{t+1} - Asset_t - C_t
\end{equation}
Where $B_t$ and $\mathbf{P_t}$ are shown in Table \ref{table_1}, $\mathbf{a_t}$ is the action given by agent at time $t$ and $\mathbf{P_t}^\prime \cdot \mathbf{W_t}$ is the inner product of vectors.

For the RL problem, the agent’s objective is to maximize its expected reward. The reward can assess the quality of each action. Furthermore, the agent optimizes the strategy under the guidance of the reward, noting the cumulative reward as $R_t$ in time $t$:
\begin{equation}
    R_t = \Sigma_{\tau=1}^t r_{\tau}
\end{equation}

\subsection{Transition Dynamics}
We start with the definition of transition probability.

$\bullet$ \textbf{Transition Probability} $\mathcal{P}$: When The change of market conditions is considered as the state transition function. $\mathcal{P}: S \times A \times S \rightarrow [0,1]$ is a function of probabilities of state transitions. $$P_{s,s'}^a = \mathcal{P}(s_{t+1} = s' | s_t = s, a_t = a)$$.

In PETS, MBPO, and M2AC, ensemble diagonal Gaussian distribution is used to predict state transition dynamics. From stochastic analysis, the state $s_t$ can be modeled by the following stochastic differential equation (SDE):
\begin{equation}
    dX_t = b(X_t,u_t)dt + \sigma(X_t,u_t)dW_t\label{SDE_control}
\end{equation}
Where $b(X_t,u_t)$ and diagonal $\sigma(X_t,u_t)$ are dirft term and difussion term. One can identify the unknown drift and diffusion terms in equation~\ref{SDE_control} through real-state data. The initial condition $X_0 = \mathbf{x} \in \mathbb{R}^d$, and $\mathbf{u_t} \in \mathbb{R}^{d}$ is an $\mathbb{F}_t$-adapted control field and $\mathbf{W_t}$ is a $d$-dimensional $\mathbb{F}_t$-standard Brownian Motion.

$\bullet$ \textbf{Policy} $\pi$: Policy is a mapping characterized by the policy $\pi :\mathbb{S} \rightarrow \mathbb{A}$. $\forall t \in {1, . . . , T}$. The aim of policy $\pi$ is to maximize its total expected portfolio value.

The agent in state $s_t \in \mathbb{S}$ takes an action $a_t \in \mathbb{A}$ which follows policy $\pi$, then receives the reward $r_t = \mathcal{R}(s_t, a_t)$ and transits to the next state $s_{t+1}$ according to the transition probability $\mathcal{P}$. The state transition dynamics in PETS, MBPO, and M2AC can be depicted as the discretized Euler form of equation~\ref{SDE_control}. The core optimization problem in RL is to find the optimal strategy $\pi^*$ that optimizes total accumulated reward.

\subsection{Policy Optimization}
Among the two main RL branches(Q-learning and policy-gradient), a lot of research has been studied to utilize the advantages of both. The actor-critic(AC) has been used for trading a large amount of stocks~\cite{xiong2018practical}. As a typical example of AC algorithms, the Deep deterministic policy gradient (DDPG) algorithm (Lillicrap et al. (2015)~\cite{lillicrap2015continuous}) shows great potential in stock trading problems and achieves some good results~\cite{emami2016deep,azhikodan2019stock}. However, it suffers from excessive sensitivity to hyper-parameters. Tuomas Haarnoja proposed a soft actor-critic (SAC) Algorithm~\cite{haarnoja2018soft} to solve the problem by employing entropy as the regularization term.

For random variable $x$ with probability density $p$, the entropy $H$ of $x$ can be defined as:
\begin{equation}
    H(p) = \underset{x\thicksim p}{\mathbb{E}}[-log p(x)]
\end{equation}
Considering discount factor $\gamma$, the entropy term is added to the expected reward, then the soft value function (soft Q function) becomes:
\begin{equation}
\begin{aligned}
    Q_{soft}^\pi(s, a)= & \underset{(s_t,a_t)\thicksim \rho_\pi}{\mathbb{E}} [\Sigma_{t = 0}^T \gamma^tr(s_t, a_t) + \\
    & \alpha \cdot \Sigma_{t=1}^T \gamma^t H(\pi(\cdot|s_t)) | s_0=s,a_0=a]
\end{aligned}
\end{equation}
Where $\rho_\pi$ represents the distribution of the state-action pair of the agent under policy $\pi$ and T is the total number of trading days.
Correspondingly, the soft V function becomes:
\begin{equation}
    V_{soft}^\pi(s) = \underset{(s_t,a_t)\thicksim \rho_\pi}{\mathbb{E}} [\Sigma_{t = 0}^T \gamma^t(r(s_t, a_t) + \alpha H(\pi(\cdot|s_t)))|s_0=s]
\end{equation}
We formulate this problem as a stochastic optimal control problem (SOC) as follows:

 \begin{equation}
 \label{soc_combine}
  \left\{
   \begin{aligned}
   V^a(x) &= \underset{a}{max}\mathbb{E}[\int_0^T r(X_s,a_s)e^{-\gamma s} ds |X_0^a = x] - \alpha log \pi(a|s)\\
   dX_t &= b(X_t,a_t)dt + \sigma(X_t,a_t)dW_t\\
   \end{aligned}
   \right.
  \end{equation}
Similar to how Proposition 3.5 in Yong's book (\cite{yong1999stochastic}) was derived, this problem's HJB equation can be presented where the solution is achieved by resolving the associated PDE.

\section{MBRL with Normalizing Flows}
\subsection{Data Analysis}
In this section, we define the alpha-stable L$\Acute{e}$vy process $L^\alpha_t$ mathematically and apply it to describe the properties of the experimental data.

\textbf{Definition 1.} A scalar random variable $X$ is called stable, if for any positive constant $a$ and $b$, the following equation holds for some $c \in \mathbb{R}^{+}$ and $b \in \mathbb{R}$
$$
a X_1+b X_2 \stackrel{d}{=} cX+d
$$
Here $X_1$, $X_2$ are i.i.d with $X$. The signal $\stackrel{d}{=}$ means equality in distribution. There is another equivalent definition of stable distribution with the characteristic function $\Phi_X(u) \triangleq \mathbb{E} e^{i \mu X}$.

\textbf{Definition 2.} A random variable X is stable if the characteristic function has the following representation:
$$
\Phi_X(u)= \begin{cases}\exp \left\{-\sigma^\alpha|u|^\alpha\left(1-i \beta(\operatorname{sign} u) \tan \frac{\pi \alpha}{2}\right)+i \mu u\right\} & \text { if } \alpha \neq 1, \\ \exp \left\{-\sigma|u|\left(1+i \beta \frac{2}{\pi}(\operatorname{sign} u) \ln |u|\right)+i \mu u\right\} & \text { if } \alpha=1 .\end{cases}
$$
Where the stability index ( tail index) $\alpha \in(0,2]$, the symmetry parameter $\beta \in[-1,1]$, the scaling parameter $\sigma > 0$, the shift parameter $\mu \in \mathbb{R}$, and $sign(u)$ is symbolic function. These four parameters determine an alpha-stable distribution $S_\alpha(\sigma, \beta, \mu)$. Note that $\Phi_X(u) = \exp\{i\mu u -\frac{1}{2}\sigma^2 u^2\}$ is the characteristic function for the Gaussian random variable.

\textbf{Definition 3.} L$\Acute{e}$vy process (Paul (1955)~\cite{Levy1955ThorieDL}), if a stochastic process $L(t), t \geq 0$, satisfies the following conditions (Duan (2015)~\cite{duan2015introduction}): \\
(i) $L_0=0$, a.s.\\
(ii) Independent increments: for $t_1<t_2<\cdots<t_{n-1}<t_n$, the random variables $L_{t_2}-L_{t_1}, \ldots, L_{t_n}-L_{t_{n-1}}$ are independent.\\
(iii) Stationary increments: $L_t-L_s$ and $L_{t-s}$ have the same distribution.\\
(iv) Stochastically continuous sample paths (i.e., sample paths are continuous in probability): for all $\delta>0$ and all $s \geq 0$,
$$
\mathbb{P}\left(\left|L_t-L_s\right|>\delta\right) \rightarrow 0
$$
as $t \rightarrow s$.

One special but important type of an L$\Acute{e}$vy process is the $\alpha$-stable L$\Acute{e}$vy process. The precise definition is as follows:

\textbf{Definition 4.} An alpha-stable L$\Acute{e}$vy process $L^\alpha_t$ is a L$\Acute{e}$vy process with an incremental distribution of $S_\alpha(\sigma, \beta, \mu)$.
We will use $S_\alpha(\sigma, \beta, \mu)$ to denote $\alpha$-stable L$\Acute{e}$vy distribution in later sections.

We estimate the parameters of the difference between two consecutive days' stock prices on the training set (04/01/2011 to 04/07/2017). The data set presented here is a 1-day dataset; further details are provided in Section 4. For each stock, the difference between two consecutive days' stock prices is denoted as $\delta \mathbf{P_t} = \mathbf{P_{t+1}}-\mathbf{P_t}$. For consistency with the data of our back-testing experiments, without any processing of the differential data.

\begin{figure}[htbp]
    \centering
	\begin{minipage}{0.45\linewidth}
		\centering
		\vspace{-0.3cm}
		\setlength{\abovecaptionskip}{0.28cm}
		\includegraphics[width=\linewidth]{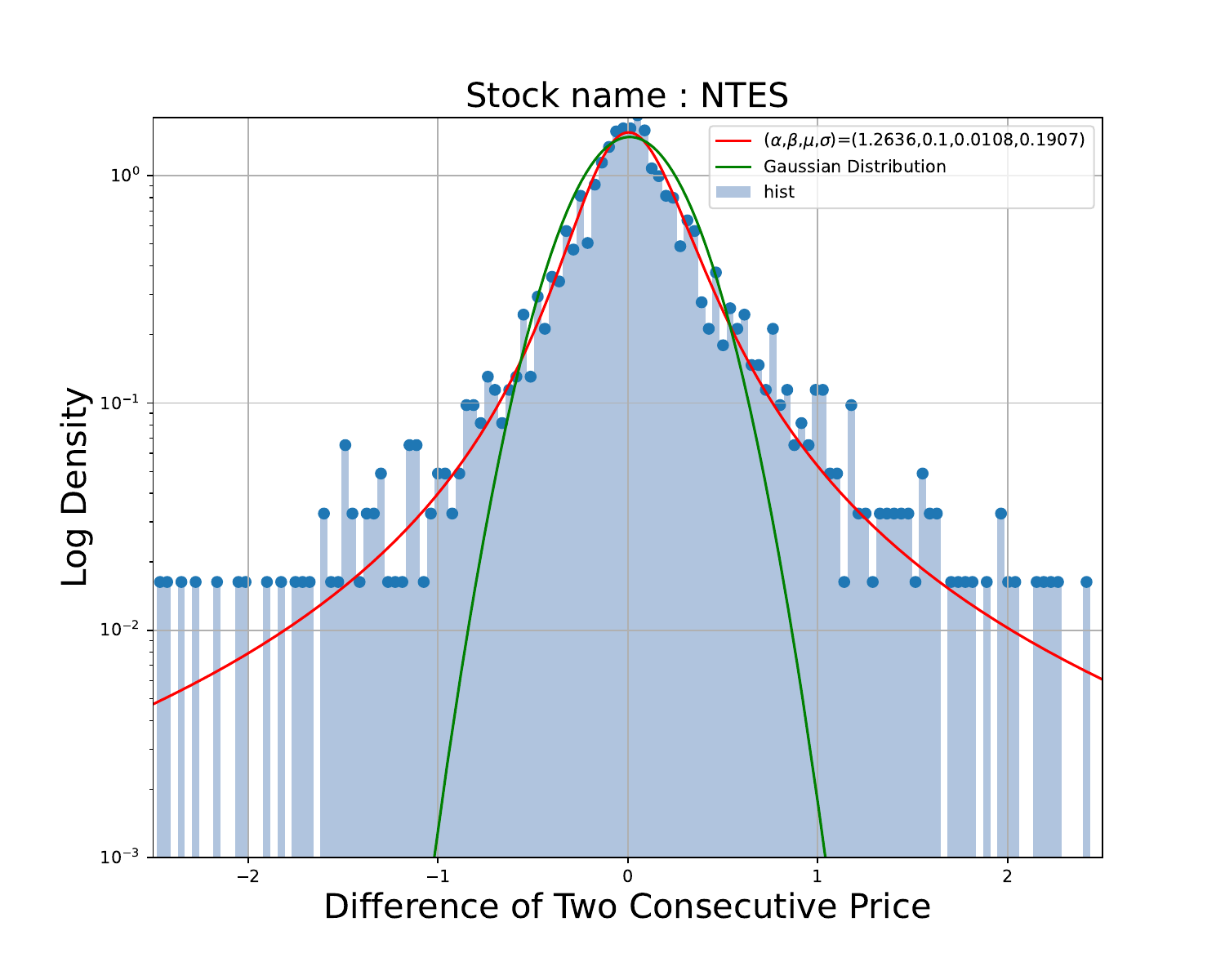}
        \captionsetup{font=small}
        \caption{Histogram of true empirical data (blue plot) and
fitted histogram from estimated $S_\alpha(\sigma, \beta, \mu)$ (red line) compared with Gaussian distribution (green line). The blue bin is the true data distribution. The X-axis is the difference between two consecutive days' stock prices, and the Y-axis is the Log form of data density.}
    \label{fit_NTES}
	\end{minipage}
	\hfill
	\begin{minipage}{0.45\linewidth}
		\centering
  \resizebox{1.0\textwidth}{!}{
        \begin{tabular}{ccccc} 
    \hline
    \multicolumn{5}{c}{NASDAQ}                           \\ 
    \hline
    ID & $\alpha$  & $\beta$    & $\mu$    & $\sigma$   \\
    \hline
    ASML       & 1.6850  & -0.0400   & 0.0627 & 0.7194  \\
    CMCSA      & 1.7211 & -0.0589 & 0.0206 & 0.1551  \\
    GOOGL      & 1.5915 & -0.0262 & 0.4345 & 3.8451  \\
    ISRG       & 1.6633 & 0.0550   & 0.1117 & 1.462   \\
    MCHP       & 1.6925 & 0.0855  & 0.0233 & 0.3287  \\
    MSFT       & 1.5845 & 0.1665  & 0.0064 & 0.2581  \\
    NTES       & 1.2636 & 0.1000   & 0.0108 & 0.1907  \\
    PEP        & 1.7376 & 0.0450   & 0.0321 & 0.3594  \\
    TCOM       & 1.4954 & -0.0456 & 0.0123 & 0.3532  \\
    TMUS       & 1.6107 & -0.0379 & 0.0324 & 0.3287  \\
    \hline
    \end{tabular}
    }
    \captionsetup{font=small}
    \captionof{table}{The estimation for the experimental stocks in the NASDAQ market. The four parameters are written by PyLevy which is a python package for the calculation of L$\Acute{e}$vy stable distributions, and supported by the Maximum Likelihood method.(https://pylevy.readthedocs.io/en/latest/index.html)}
    \label{NAS_fitting}
    \end{minipage}
\end{figure}

The conventional MBPO model assumes that the state has a Gaussian distribution; however, considering $\alpha$-stable L$\Acute{e}$vy distribution, we obtain different outcomes. Figure \ref{fit_NTES} shows the fitted histogram using the $S_\alpha(\sigma, \beta, \mu)$ distribution for NTES as an example. The estimated distribution well characterizes the fat-tail behaviors and the bulk portion of price's difference distributions. In Table \ref{NAS_fitting}, we list the four parameters obtained by estimating the $S_\alpha(\sigma, \beta, \mu)$ distribution on the experimental stocks data of the NASDAQ market. Note that the parameters estimated of the Dow market and S\&P market are shown in \ref{DOW_fitting_appendix} and \ref{SP_fitting_appendix} in Appendix A, respectively. 

 
 It is worth mentioning that the tail parameter $\alpha$ is typically between 1.5 and 1.7, although our data may not have been perfectly fitted using the $L\Acute{e}vy$ distribution, they do consistently exhibit the heavy-tailed property. As it is known that, when $\alpha$ equals  2, the distribution becomes Gaussian, which suggests that the experimental data may be somewhat close to the Gaussian distribution. This explains why in our experiments, good experimental performance can be obtained using the conventional diagonal Gaussian model. The majority of the calculated parameters $\beta$ and $\mu$ are close to zero, which implies that the spreads are not so skewed. These findings provide additional evidence that price fluctuations in the stock market exhibit symmetry property. The four parameters $\alpha$, $\beta$, $\mu$, and $\sigma$, can offer clues to illustrate the complexity properties of empirical data. In this paper, we aim to simulate the high-dimensional joint probability of the complex trading environment to better understand the intrinsic mechanisms of quantitative trading. Now, we introduce the heavy-tailed normalizing flows.


\subsection{Heavy-tailed Normalizing Flows}
In this part, we introduce the normalizing flows model and describe two heavy-tailed distributions, then define the heavy-tailed normalizing flows model.

\subsubsection{Normalizing Flows}
Normalizing flows~\cite{dinh2016density} is a compound transformed function which consists of a series of bijective transformations. It can generally be expressed as:
$$\mathbf{x} = F_\theta(\mathbf{z}) = f_n \circ f_{n-1} \circ f_{n-2} \circ \cdots \circ f_{2} \circ f_{1} (\mathbf{z})$$
Where $\theta$ denotes the neural network parameters, n is the transformation steps, and $\mathbf{z}$ follows a sample distribution (e.g. normal) which is the base distribution with density function $p_{\mathbf{z}}(\mathbf{z})$. $\mathbf{x}$ represents the unknown input data with density function $p_{\mathbf{x}}(\mathbf{x})$.
Employing the change of variables formula, we can derive $p_\theta(\mathbf{x})$ as
\begin{equation}
    p_\theta(\mathbf{x})=p_{\mathbf{z}}(F^{-1}_\theta(\mathbf{x}))\left |det\left(\frac{\partial F^{-1}_\theta(\mathbf{x})}{\partial \mathbf{x}}\right)\right |
\end{equation}
Where $\frac{\partial F^{-1}_\theta}{\partial \mathbf{x}}$ is the Jacobian matrix, and $p_\theta(\mathbf{x})$ is the estimated density under given observed data $\mathbf{x}$ with transformation neural network $F_\theta$.

A simple but effective improved version of the flow-based algorithm is Real-NVP (Dinh (2016)~\cite{dinh2016density}) which combines addition and multiplication into the affine layer and sets the simple function $f_{i}$ as follows:
$$
    \begin{cases}
        \mathbf{z}^{1:d}=\mathbf{x}^{1:d}\\
    \mathbf{z}^{d+1:D}=\mathbf{x}^{d+1:D}\odot exp(s_\phi(\mathbf{x}^{1:d}))+t_\varphi(\mathbf{x^{1:d}})
    \end{cases}
$$
Where $\odot$ is the element-wise product, $\phi$ and $\varphi$ is the neural network parameters. $s_\phi(\cdot)$ and $t_\varphi(\cdot)$ are called scaling and translation functions respectively which map $\mathbb{R}^d\rightarrow \mathbb{R}^{D-d}$. In this paper, we employ Real-NVP as the flow-based model and we shall continue to write it as a normalizing flows model for writing convenience.




Updating the model parameters $\theta$ can be done via maximum likelihood technology. Given batch data $D$, the algorithm's aim is to maximize the average log-likelihood given by:
\begin{align}
    \mathcal{L}(x;\theta)=\frac{1}{|D|}\sum\limits_{\mathbf{x}\in D}log\ p_\theta(\mathbf{x}) \label{likelihood_origin}
\end{align}
Where $|D|$ denotes the sample number of batch data $D$.

Although it appears that NF may be transferred from any type of base density $p_{\mathbf{z}}(\mathbf{z})$ to the unknown density $p_\mathbf{x}(\mathbf{x})$, it has been demonstrated that an invertible transformation of the flow-based model cannot change it from a light-tailed distribution to a heavy-tailed distribution (Jaini (2020)~\cite{jaini2020tails}, Huster (2021)~\cite{huster2021pareto}). For the experimental data in this paper, we have numerically estimated their $S_\alpha(\sigma, \beta, \mu)$ distribution parameters to prove that they obey some sort of heavy-tailed distribution. Hence, as base distributions, we'll provide two popular heavy-tailed distributions.

\subsubsection{Heavy-tailed Distribution}
The high-dimensional density function of the generalized Gaussian distribution and the $t$-distribution will be covered in this subsection.

\textbf{Definition 5.} Generalized Gaussian Distribution (GGD) 
The GGD is a continuous probability distribution with probability density function given by:
$$f_{GGD}(y; \alpha, \beta, \mu) = \frac{\beta}{2 \alpha \Gamma(1/\beta)} \exp\left[-\left(\frac{|y - \mu|}{\alpha}\right)^\beta\right]$$
Here, the shape parameter $\beta > 0$, the scale parameter $\alpha > 0$, the location parameter $\mu \in \mathbb{R}$ and $\Gamma(\cdot)$ is the gamma function. When $\beta = 2$, the GGD reduces to the normal distribution.
The maximum likelihood \ref{likelihood_origin} will be rewritten as:
\begin{align}
    \mathcal{L}_{GGD}(x;\theta)=\frac{1}{|D|}\sum\limits_{\mathbf{x}\in D}log \left(\frac{\beta}{2 \alpha \Gamma(1/\beta)} \exp\left[-\left(\frac{|F^{-1}_\theta(\mathbf{x}) - \mu|}{\alpha}\right)^\beta\right] \left| det\left(\frac{\partial F^{-1}_\theta(\mathbf{x})}{\partial \mathbf{x}}\right)\right|\right)
\end{align}

\textbf{Definition 6.} The $t$-distribution (Rice (2006)~\cite{rice2006mathematical})
If $Z \sim N(0,1)$ and $U \sim \chi_n^2$ and $Z$ and $U$ are independent, and the distribution of $Z / \sqrt{U / n}$ is called the $t$-distribution with $n$ degrees of freedom.
The density function of the $t$-distribution with $n$ degrees of freedom is
$$
f_t(y; n)=\frac{\Gamma[(n+1) / 2]}{\sqrt{n \pi} \Gamma(n / 2)}\left(1+\frac{y^2}{n}\right)^{-(n+1) / 2}
$$
Where $\Gamma(\cdot)$ is the gamma function. The $t$-distribution (n=1) is the Cauchy distribution and the $t$-distribution ($n \rightarrow \infty$) is towards the normal distribution.
The maximum likelihood \ref{likelihood_origin} can be rewritten as such:
\begin{align}
    \mathcal{L}_t(x;\theta)=\frac{1}{|D|}\sum\limits_{\mathbf{x}\in D}log \left(\frac{\Gamma[(n+1) / 2]}{\sqrt{n \pi} \Gamma(n / 2)}\left(1+\frac{F^{-1}_\theta(\mathbf{x})^2}{n}\right)^{-(n+1) / 2} \left| det\left(\frac{\partial F^{-1}_\theta(\mathbf{x})}{\partial \mathbf{x}}\right)\right| \right)
\end{align}

\subsubsection{Heavy-tailed Normalizing Flows}
We denote the normalizing flows model with the heavy-tailed base distribution as the heavy-tailed NF (HNF) model. Employing the GGD distribution and $t$-distribution as the base distributions, we use heavy-tailed normalizing flows model to approximate the joint distribution of high-dimensional non-Gaussian state features.

\begin{figure}[htbp]
    \centering
    \subfigure[$t$-distribution (n = 8)]{
    \label{NTES_t}
    \includegraphics[width=6cm]{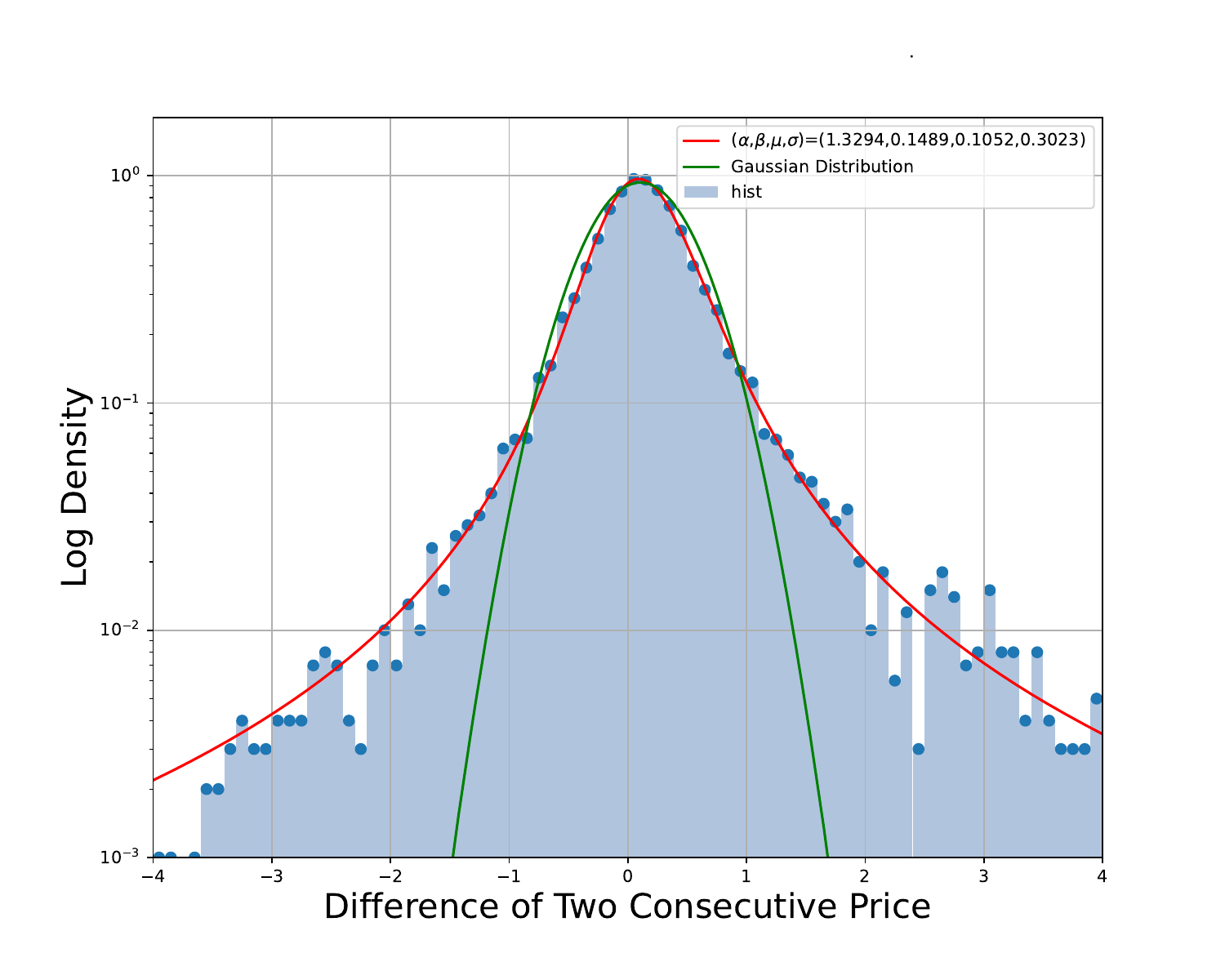}}
    \subfigure[GGD ($\beta = 1.05, \alpha = 1$)]{
    \centering
    \label{NTES_gn}
    \includegraphics[width=6cm]{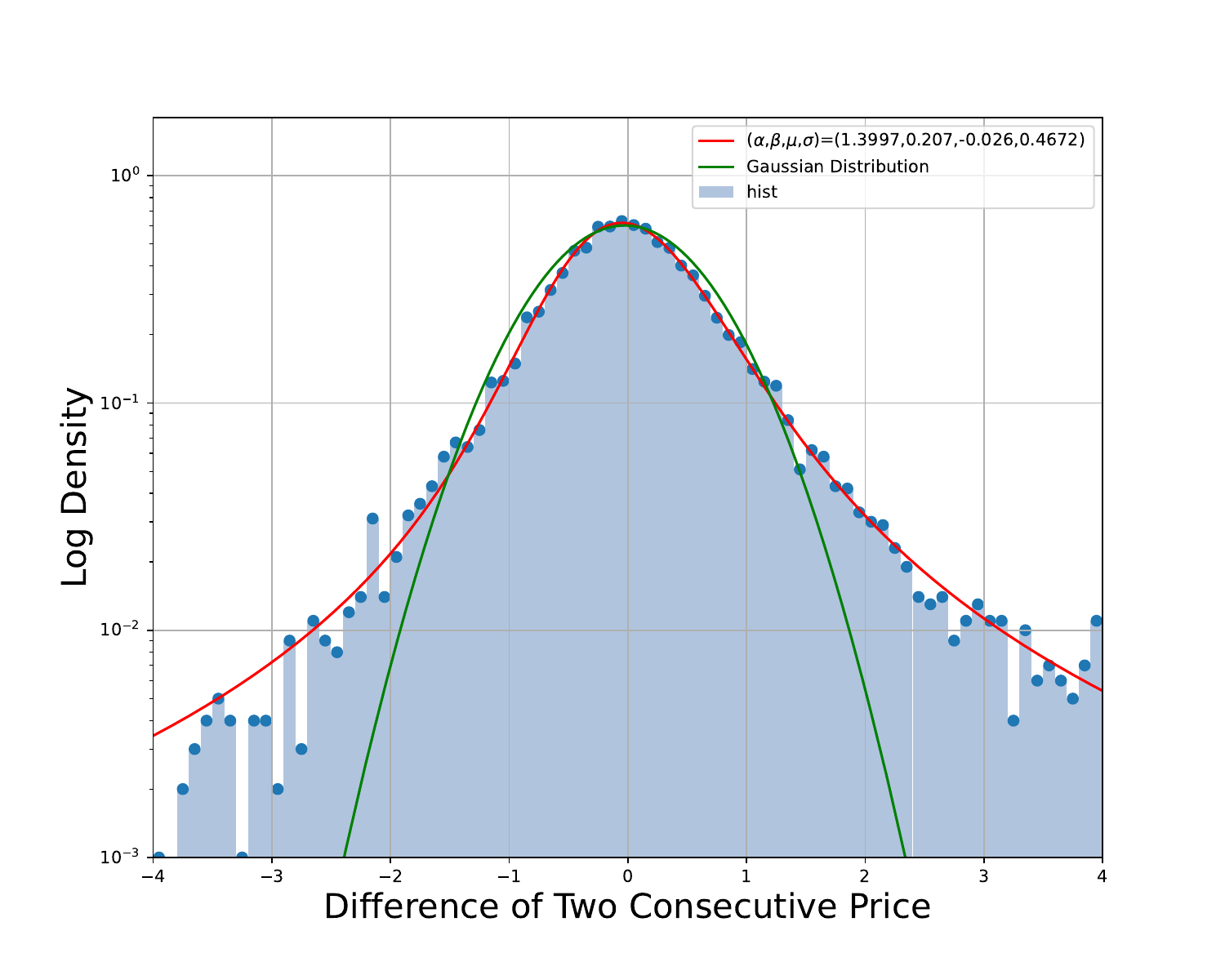}}
    \captionsetup{font=small}
    \caption{Density fitted with sample data generated by HNF. Subfigure (i) corresponds to $t$-distribution as the base distribution, Subfigure (ii) generalized Gaussian distribution. Histogram of generated data (blue plot) and fitted histogram from estimated stable distribution (red line) compared with Gaussian distribution (green line).}
    \label{heavy_tail_fitting_figure}
\end{figure}

As for verification in Figure \ref{heavy_tail_fitting_figure}, we compare our approximated density projected to one stock (NTES) with $t$-distribution and generalized Gaussian distribution, respectively. Histograms are plotted with 10000 sampled data from the approximated distribution. When fitting a smooth curve with the histograms, we obtain the values of four parameters. Remembering from Section 3.1 that the four parameters fitted by real data is (1.2636, 0.1 0.0108 0.1907). When $t$-distribution is used, the estimated tail parameter of the generated data differ from the true tail parameter by just 0.0658 (1.3294-1.2636) and the maximum error value from the true parameters is 0.1116 (0.3023-0.1907). The fitted shift parameter of the generalized Gaussian distribution deviate from the real parameter by only 0.01, and the maximum error value from the true parameters is 0.2756 (0.4672-0.1907). This implies that the NF can learn to characterize the distribution of the experimental data using our heavy-tailed underlying distribution.

\subsection{Proposed Model}
Based on the heavy-tailed NF model as explained in the previous session, we know it is worthwhile to simulate the high-dimensional joint probability of the complex trading environment. In \cite{lu2021learning}, the authors establish a link between temporal normalizing flows of an SDE and the corresponding solution. In our case, the density $\mathcal{P}_x(x)$ can be treated as the solution of the Fokker-Planck equation corresponding to equation~\ref{SDE_control}. This inspires us to simulate state transition dynamics through the heavy-tailed NF model.

Our goal here is to predict the next state $s_{t+1}$ from the current state $s_t$ and action $a_t$. From the perspective of optimal control, we discretized the transition dynamical system in terms of the Euler-Maruyama scheme:
\begin{equation}
s_{n+1} - s_n = b(s_n, a_n)\Delta t + \sigma (s_n,a_n)d L^\alpha_t, \quad \quad n = 0,1,2,\cdots, T
\end{equation}
In our case, the term $b(s_n, a_n)$ is zero. With real high-dimensional time-series data as our states, we learn the density of the stationary increment ${s_{t+1} - s_t}$. Afterward, given the current state $s_t$, the next state $s_{t+1}$ is obtained by adding $s_t$ with samples generated from the approximated density.

From the last session, we know that our main task is to simulate the probability density function $\mathcal{P}_X$ through heavy-tailed NF by solving the equation \ref{soc_combine}. In fact, the $\mathcal{P}_X$ can help us to learn more information about equation \ref{SDE_control}. One way to study this information is through constructing an inverse problem and learning unknown drift and diffusion terms in equation \ref{SDE_control} with the method of maximum log-likelihood~\cite{fang2022end}. In this way, we could learn the intrinsic hidden dynamics from real data as a stochastic dynamical system, thereby establishing a link between optimal control and reinforcement learning. Once the unknown drift and diffusion coefficients are obtained, the RL problem could be addressed by solving an HJB equation, under some conditions, which is beyond the scope of our discussions in this paper.

The MBNF algorithm's pseudo-code is listed in Table \ref{table_19}.





\begin{table*}[!ht]
\renewcommand{\arraystretch}{1.25}
\caption{MBNF Framework}
\label{table_19}
\centering
\begin{tabular}{l}
\toprule
\textbf{Algorithm of MBNF: MBPO with heavy-tailed normalizing flows }\\
\midrule
Initialize policy $\pi_\theta$, heavy-tailed normalizing flows model $\Phi_{HNF}$, real distribution $\mathcal{P}(x)$\\
real environment buffer $\mathcal{B}_{env}$ and agent training buffer $\mathcal{B}_{agent}$\\
\bf{for 1:E steps do}\\
\quad  $\mathcal{B}_{env}$ goes one step further and adds the data to itself\\
\quad \bf{for every M steps do} \\
\quad \quad Train heavy-tailed normalizing flows model $\Phi_{HNF}$ on $\mathcal{B}_{env}$ :\\
\quad \quad \bf{for 1:N steps do}\\
\quad \quad \quad $\Phi_{HNF}$ computes approximate distribution $\widetilde{\mathcal{P}}$ through Maximum Likelihood Function: \\
\quad \quad \quad${\max}_{\theta} \mathcal{L}_t(x;\theta)$ \\
\quad \quad Sample $x \thicksim \widetilde{\mathcal{P}}$ and update agent training buffer $\mathcal{B}_{agent}$ :\\
\quad \quad Select a batch of data from $\mathcal{B}_{env}$ as current states $s_{t}$\\
\quad \quad Get a batch of action $a_t$ using policy $\pi_{\theta}$ \\
\quad \quad Get a batch of predicted states $\widetilde{s}_{t+1}$ and predicted $\widetilde{r}_{t+1}$ with trained $\Phi_{HNF}$ using $a_t$ \\
\quad \quad Add $s_{t}$, $a_t$, $\widetilde{s}_{t+1}$, $\widetilde{r}_{t+1}$ to $\mathcal{B}_{agent}$ \\
\quad \bf{for 1:L steps do} \\
\quad \quad Get a batch of quadruplet $s_{t}$, $a_t$, $\widetilde{s}_{t+1}$, $\widetilde{r}_{t+1}$ from $\mathcal{B}_{agent}$ \\
\quad \quad Update policy parameters on $\mathcal{B}_{agent}$: $\theta\:\leftarrow\:\theta\:-\:\lambda_\pi\:\triangledown_\theta\:(\mathcal{J}_\pi(\theta, \mathcal{B}_{agent}))$\\
\bottomrule
\end{tabular}
\end{table*}

\section{Experiment}
In this section, we present some experimental results and corresponding explanations, from which we can see how and why heavy-tailed NF help to improve the performance of the classical model-based reinforcement learning model in stocks portfolio.

\subsection{Dataset}
All of our sample data is from the Yahoo finance database.
To compare the effectiveness of our strategies across different markets, we conducted our experiments on three markets: the Dow Jones Industrial Average (Dow), the National Association of Securities Dealers Automated Quotations (NASDAQ) and the Standard and Poor's 500 (S\&P). The Dow represents the trend of traditional U.S. industrial companies, while the NASDAQ favors technology companies and emerging sectors; The S\&P is the most comprehensive of the three markets.

Considering all available stocks during the whole experiment period, 10 stocks are selected by their turnover rate within 180 days before Jan. 04, 2011 in the market. The selected stocks have relatively low turnover rates so as to avoid big noise or fluctuations in unstable financial markets with small market capitalization. Here, as a liquidity indicator, the "turnover rate" refers to the frequency of stock changing hands in the market within a certain period of time. In this way, we can better explore the applicability and generalization capability of the proposed algorithm in this paper.

For each experiment, the dataset has a historical daily price data from Jan. 4, 2011 to Mar. 31, 2021. Data from Jan. 4, 2011 to Jul. 3, 2017 (\textbf{1635 days}) was used as the training set, and data from Jul. 4, 2017 to Jan. 30, 2019 (\textbf{396 days}) was used as the validation set. The remaining data from Jan. 31, 2019 to Mar. 31, 2021 (\textbf{545 days}) was used as a testing set. We trained our agent on training data, then selected model hyper-parameters by evaluating metrics such as annualized return, Sharpe-ratio, and maximum drawdown on validation data, and finally applied the selected model to the testing data, which is shown in the following figures and tables. Notice that all the testing results are averaged from 10 random experiments. 

Here we set up the following hyper-parameters for our experiments: initial balance $B_0 = 1e6$ dollars, and maximum shares per transaction $h_{max} = 100$. 

\subsection{Portfolio Performance}
The performance of MBNF and MBPO strategies was compared in the three markets. Both strategies for testing data were compared with the Market Index baselines: DJIA, NDX and GSPC respectively.

For the evaluation metrics, we used the following measurements to evaluate the proposed model performance. 
\begin{itemize}
    \item{Annualized Return: the geometric average amount of money earned by an investment strategy each year over a given time period.}
    \item{Cumulative Return: the overall effect of a trading strategy in a certain time range.}
    \item{Annualized Volatility: the annualized standard deviation of portfolio return which shows the robustness of the agent.}
    \item{Sharpe-ratio~\cite{sharpe1998sharpe}: the return earned per unit volatility which is a widely-used measure of the performance in investment.}
    \item{Calmar Ratio: the annualized return earned per unit maximum drawdown in the period.}
    \item{Stability: the index which determines the R-squared linear fit to the cumulative log returns.}
    \item{Maximum Drawdown: the maximum loss from a peak-to-trough decline of an investment before a new peak is attained.}
\end{itemize}

\noindent\textbf{Overall performance with different measurements.} In Table \ref{totle_return_table}, we evaluate the performance of MBNF algorithms by seven metrics. Despite the significant disturbance caused by COVID-19, our proposed model MBNF exhibits superior performance on all seven metrics in three markets. In the Dow market, both MBNF and MBPO methods achieve the best performance. For instance, MBNF doubles the Market Index baseline Sharpe-ratio and outperforms the Market Index baseline's annualized return, Sharpe-ratio, and stability metrics by 10\%, 0.34, and 64\%, respectively. On the other hand, the MBPO model outperforms the Market Index baseline's annualized return and Sharpe-ratio metrics by 4\% and 0.17. In the S\&P market, the MBNF model gets a Sharpe-ratio of around 1.03. However, in the NASDAQ market, both models perform relatively poorly, with the Market Index baseline scoring the highest in the maximum drawdown metrics. Our proposed model's annualized return and Sharpe-ratio index are only slightly higher than the Market Index baseline, whereas the MBPO model fails to meet the benchmark. Additionally, the maximum drawdown metrics indicate that the heavy-tailed NF can assist the agent in making more stable investment decisions with less risk. Finally, considering the scores of the Dow and NASDAQ, the S\&P market's performance reflects their comprehensiveness, with moderate scores in annualized return, Sharpe-ratio, stability, maximum drawdown, and other indicators.

Considering both risk management and profit gain, the Sharpe-ratio and Calmar ratio shows that MBNF is significantly better than MBPO model and Market Index baseline's performance. This indicates combining heavy-tailed NF with some classical reinforcement learning algorithms is a promising direction to explore.\\

\begin{table*}
\setlength{\tabcolsep}{3pt}
\centering
\caption{Performance comparison in three markets}
\label{totle_return_table}
\scalebox{0.8}{
\begin{tabular}{c|ccc|ccc|ccc} 
\hline
                      & \multicolumn{3}{c|}{Dow}                & \multicolumn{3}{c|}{NASDAQ}                   & \multicolumn{3}{c}{S\&P}                   \\ 
\hline
Back-testing Indicators   & MBNF              & MBPO     & Baseline & MBNF              & MBPO     & Baseline          & MBNF              & MBPO     & Baseline  \\ 
\hline
Annualized Return     & \textbf{23.30\%}  & 17.58\%  & 14.00\%  & \textbf{35.35\%}  & 30.26\%  & 33.88\%           & \textbf{26.16\%}  & 23.93\%  & 19.48\%   \\
Cumulative Return     & \textbf{57.31\%}  & 41.95\%  & 32.69\%  & \textbf{92.43\%}  & 77.15\%  & 87.72\%           & \textbf{65.31\%}  & 59.04\%  & 46.85\%   \\
Annualized Volatility & \textbf{25.17\%}  & 24.02\%  & 26.59\%  & \textbf{28.66\%}  & 27.30\%  & 27.96\%           & \textbf{25.88\%}  & 25.22\%  & 25.25\%   \\
Sharpe Ratio          & \textbf{0.96}     & 0.80     & 0.63     & \textbf{1.20}     & 1.11     & 1.19              & \textbf{1.03}     & 0.98     & 0.83      \\
Calmar Ratio          & \textbf{78.32\%}  & 54.64\%  & 37.75\%  & \textbf{123.75\%} & 101.33\% & 120.84\%          & \textbf{91.52\%}  & 79.72\%  & 57.43\%   \\
Stability             & \textbf{90.33\%}  & 80.04\%  & 26.05\%  & \textbf{93.30\%}  & 85.59\%  & 89.53\%           & \textbf{83.22\%}  & 78.98\%  & 63.14\%   \\
Maximum Drawdown      & \textbf{-29.76\%} & -32.18\% & -37.09\% & -28.56\%          & -29.87\% & \textbf{-28.03\%} & \textbf{-28.59\%} & -30.01\% & -33.93\%  \\
\hline
\end{tabular}
}
\end{table*}



\begin{figure}[t]
	\centering
	\vspace{-0.15in}
	\begin{minipage}{1\linewidth}	
		\subfigure[Dow Market]{
			\label{returns_dow}
			\includegraphics[width=0.45\linewidth,height=1.2in]{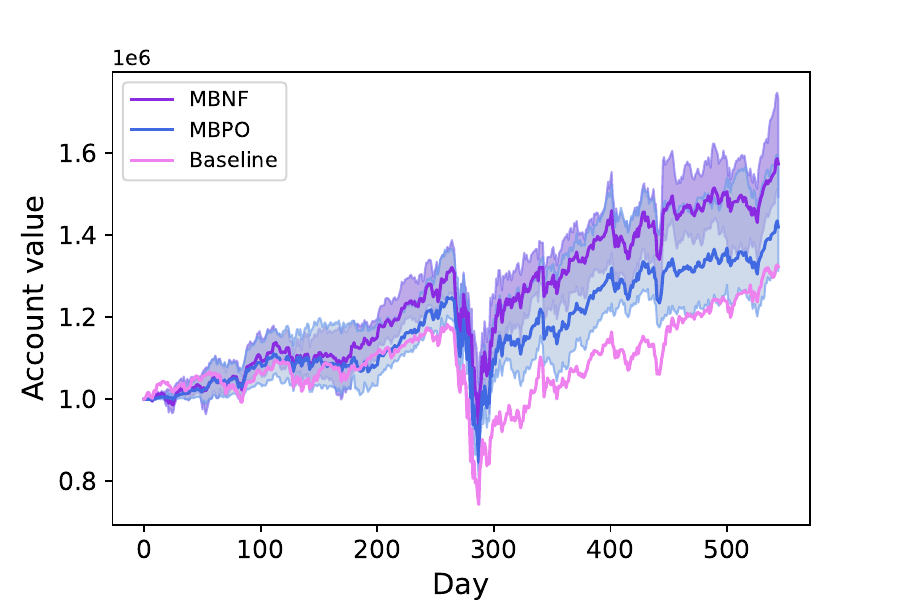}	
		}\noindent
		\subfigure[NASDAQ Market]{
			\label{returns_nas}
			\includegraphics[width=0.45\linewidth,height=1.2in]{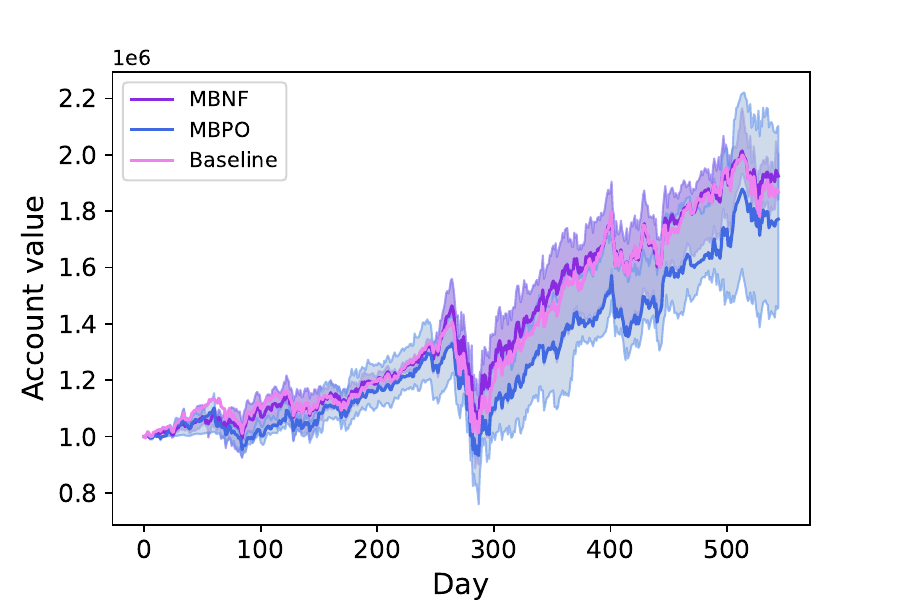}
		}
	\end{minipage}
	\vskip -0.3cm 
	\begin{minipage}{1\linewidth }
		\subfigure[S$\&$P Market]{
			\label{returns_sp}
			\includegraphics[width=0.45\linewidth,height=1.2in]{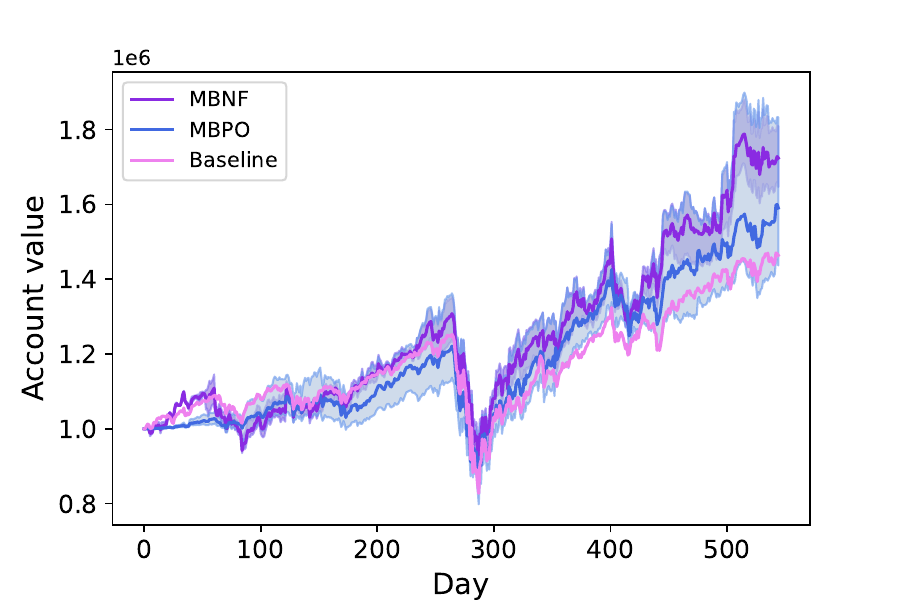}
		}\noindent
		\subfigure[Technical indicators analysis in S$\&$P market]{
			\label{cost_true_false}
			\includegraphics[width=0.45\linewidth,height=1.2in]{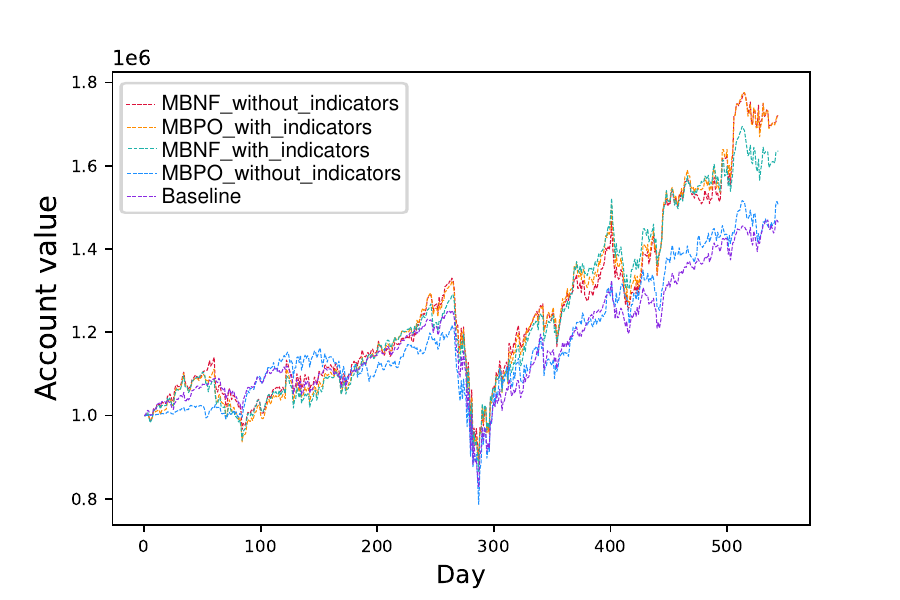}
		}
	\end{minipage}
	\vspace{-0.18in}
    \captionsetup{font=small}
	\caption{Daily return performance. Subfigures (i) (ii) (iii) correspond to the back-testing results of the daily return of the experimental stocks in the three markets of Dow, NASDAQ, and S\&P markets, respectively. The Y-axis shows the cumulative returns as of the day, and the X-axis the number of days. The MBNF (solid purple), MBPO (solid blue), and baseline (solid pink) were computed by an average of 10 independent results. The shaded area represents the range of fluctuations between the maximum and minimum values in the ten experiments. (iv) represents the effect of the use of technical indicators and cost on the MBPO and MBNF algorithms in the S$\&$P market. Note that the dotted lines are shown from the average of 10 independent experiments, and we do not provide shaded plots to allow for a clearer view of the lines.}
	\vspace{-0.2in}	
	\label{daily return performance}
\end{figure}

\noindent\textbf{Comparison of daily return.} Figure~\ref{daily return performance}\ref{returns_dow},~\ref{returns_nas} and ~\ref{returns_sp} show the back-testing outcomes for the daily cumulative returns of the two models across three different markets. The proposed model MBNF performs exceptionally well in both cases, particularly in the Dow market. The COVID-19 outbreak caused a significant reduction in the maximum drawdown measure for all models between days 280 and 300. However, in the post-epidemic bull market, our MBNF model was able to take advantage of the opportunity and generate higher profits at a faster pace. Although the proposed model cannot predict future prices, it increases its resilience by taking opportunities during a bull market. Furthermore, the blue-shaded band representing MBPO has a more volatile return curve than the purple-shaded band representing MBNF, demonstrating that MBNF is more robust than MBPO.\\

\noindent\textbf{Technical indicators for heavy-tailed normalizing flows.}
In previous experiments, we included various technical indicators in the state space to better simulate the market environment. To further analyze the robustness of our method, we tested the performance of MBNF and MBPO with and without technical indicators in the comprehensive S\&P market. As shown in Figure~\ref{daily return performance}~\ref{cost_true_false}, the performance of MBNF with and without technical indicators was not significantly different, while the MBPO model performed much better when technical indicators were used. This suggests that heavy-tailed NF can discover the relationship between input data and make good decisions without relying heavily on technical indicators, whereas the diagonal Gaussian model depends more on indicator-based trading. In subsection 4.3.1, we further analyzed the data and explained the experimental results using the pattern causality method.

\subsection{Performance Explanation}
In this section, we investigate the causal relationships between stocks and the behavior of the dynamic model during the back-testing process, providing a quantitative explanation for the experimental performance and robustness of the proposed model.

\subsubsection{Causual Analysis}

Financial markets have been extensively analyzed as complex systems, with asset pricing serving as the cornerstone for analyzing structures such as financial networks \cite{Normalizing_flows_2010contagion}
\cite{nf_boginski2005statistical}. Understanding the nature of asset price interactions is crucial for developing effective investment strategies or micro-trading tactics \cite{nf_fabozzi2008portfolio}.

This paper uses the pattern causality (PC) method of Stavroglou (2019)~\cite{nf_stavroglou2019hidden} to confirm the presence of correlations among various stocks within the same market, which is found to be true for all three markets analyzed.



\begin{figure}[ht]
	\centering
    \subfigure[Positive relationships]{
	\begin{minipage}{0.49\linewidth}
        \includegraphics[width=0.3\textwidth]{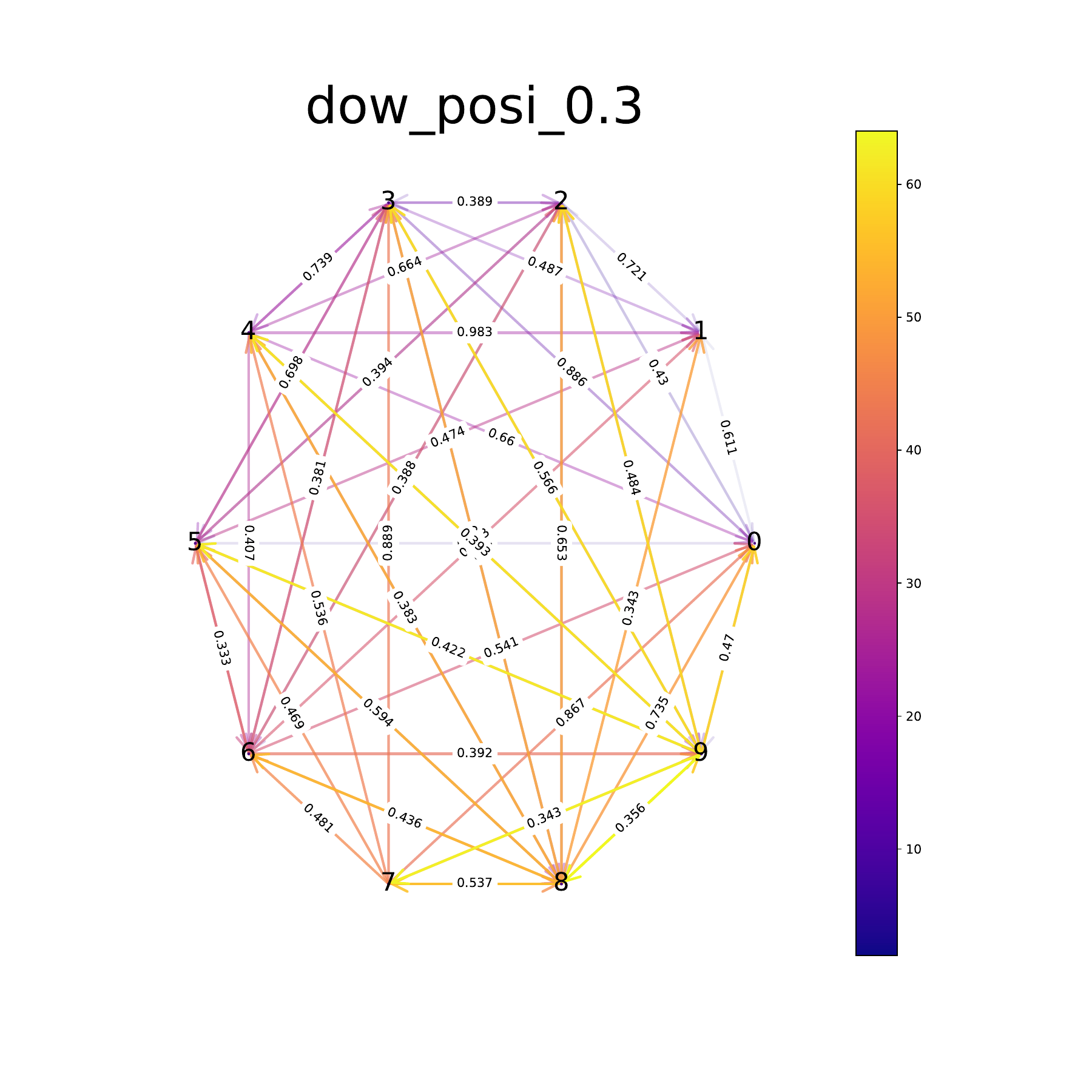}
        \includegraphics[width=0.3\textwidth]{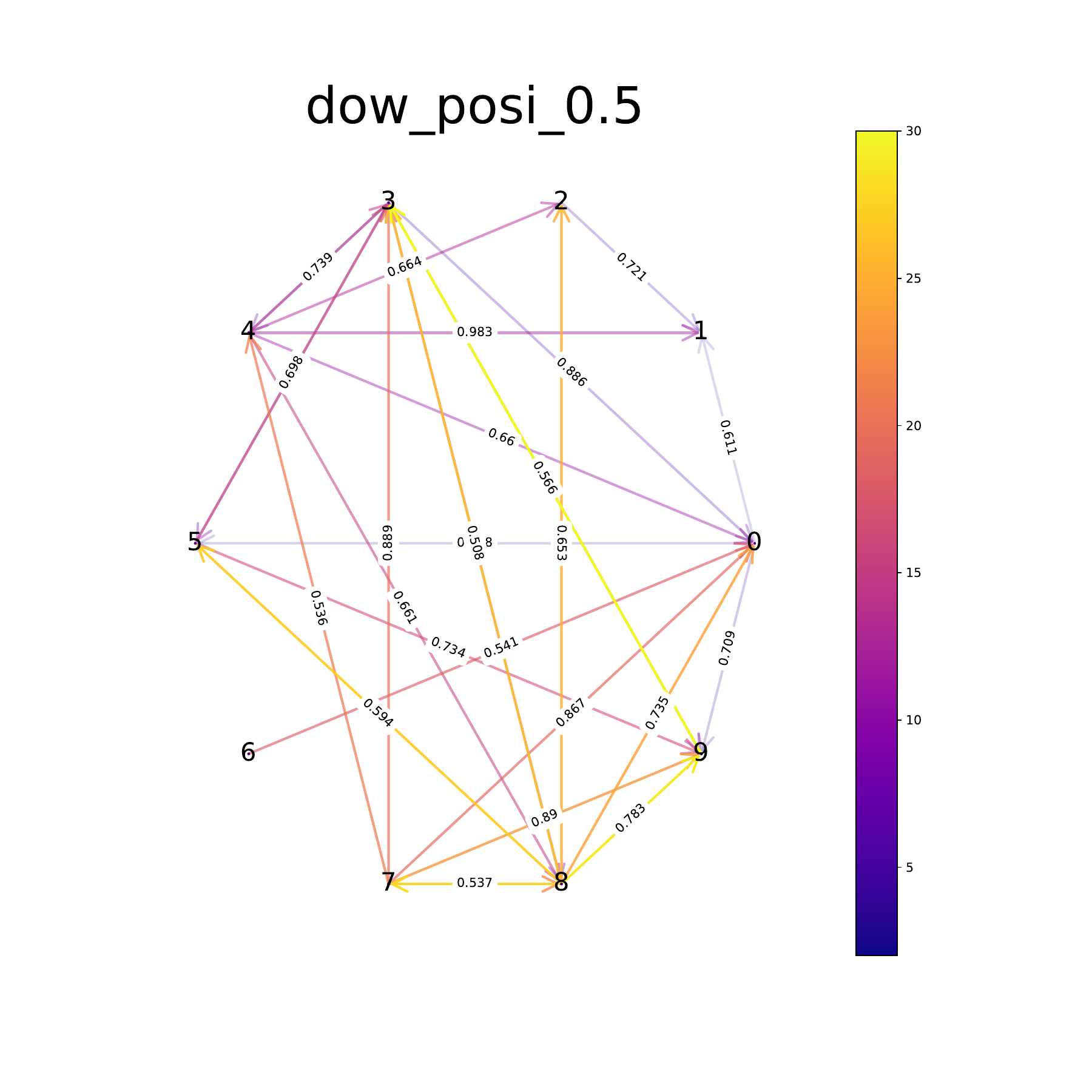}
        \includegraphics[width=0.3\textwidth]{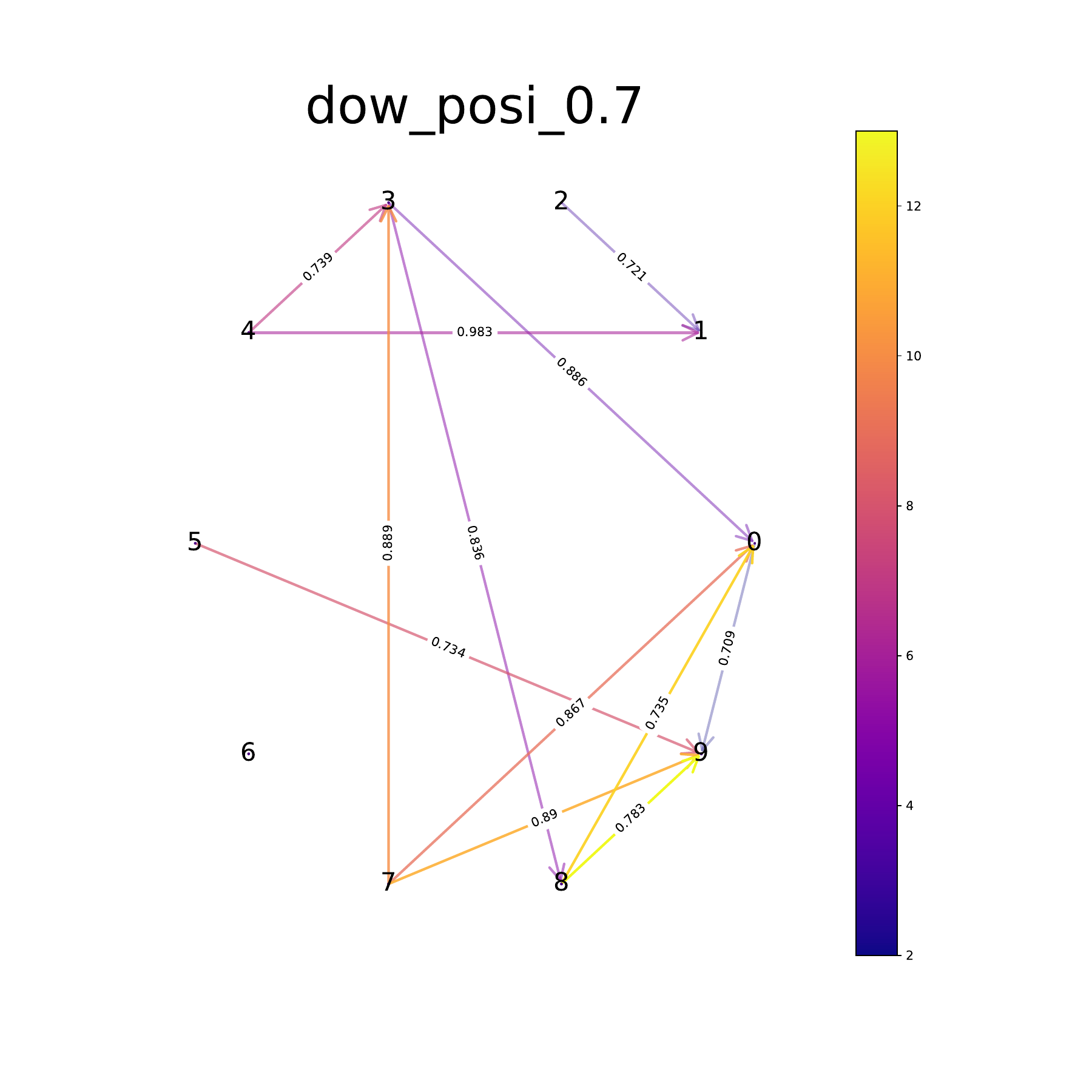}\\
        \includegraphics[width=0.3\textwidth]{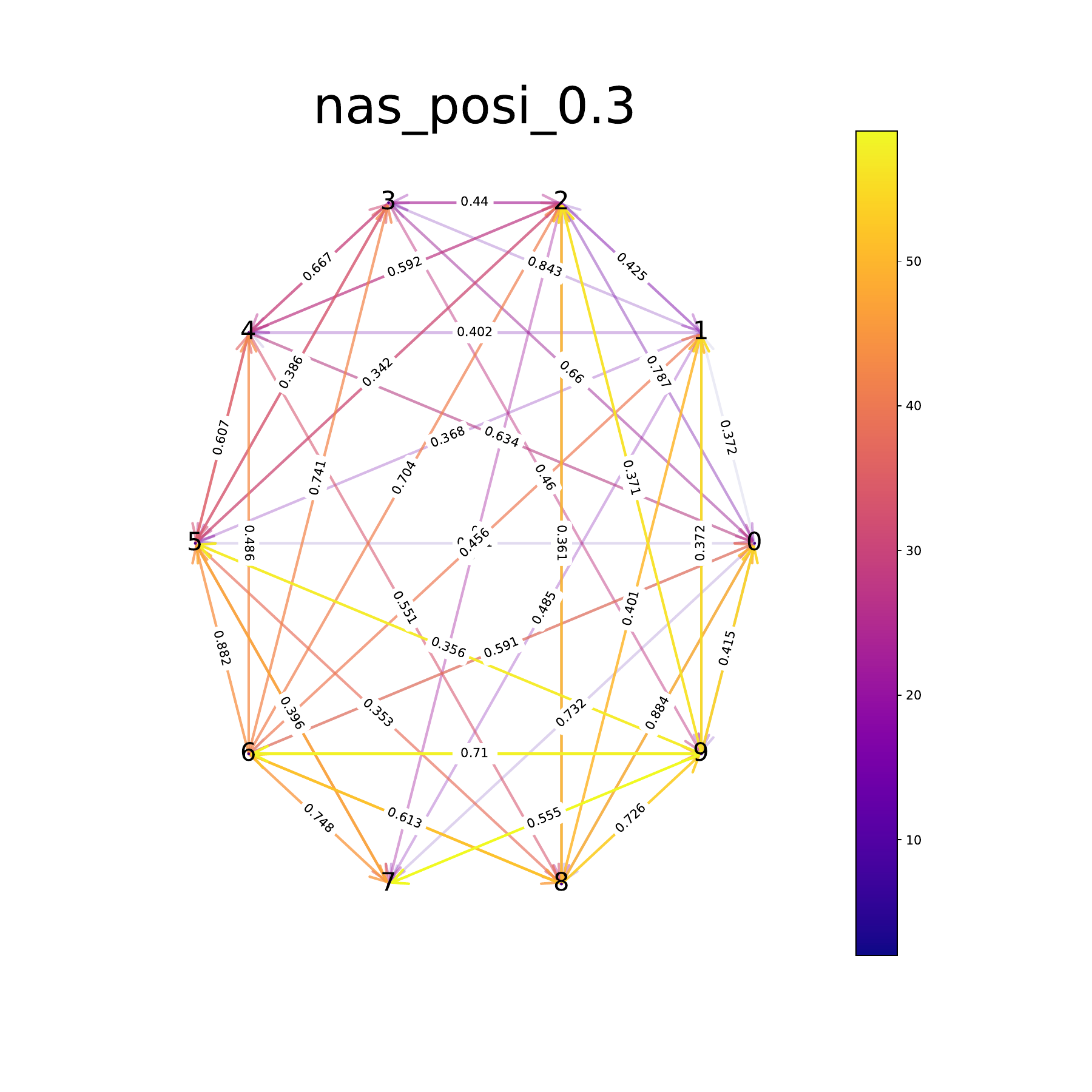}
        \includegraphics[width=0.3\textwidth]{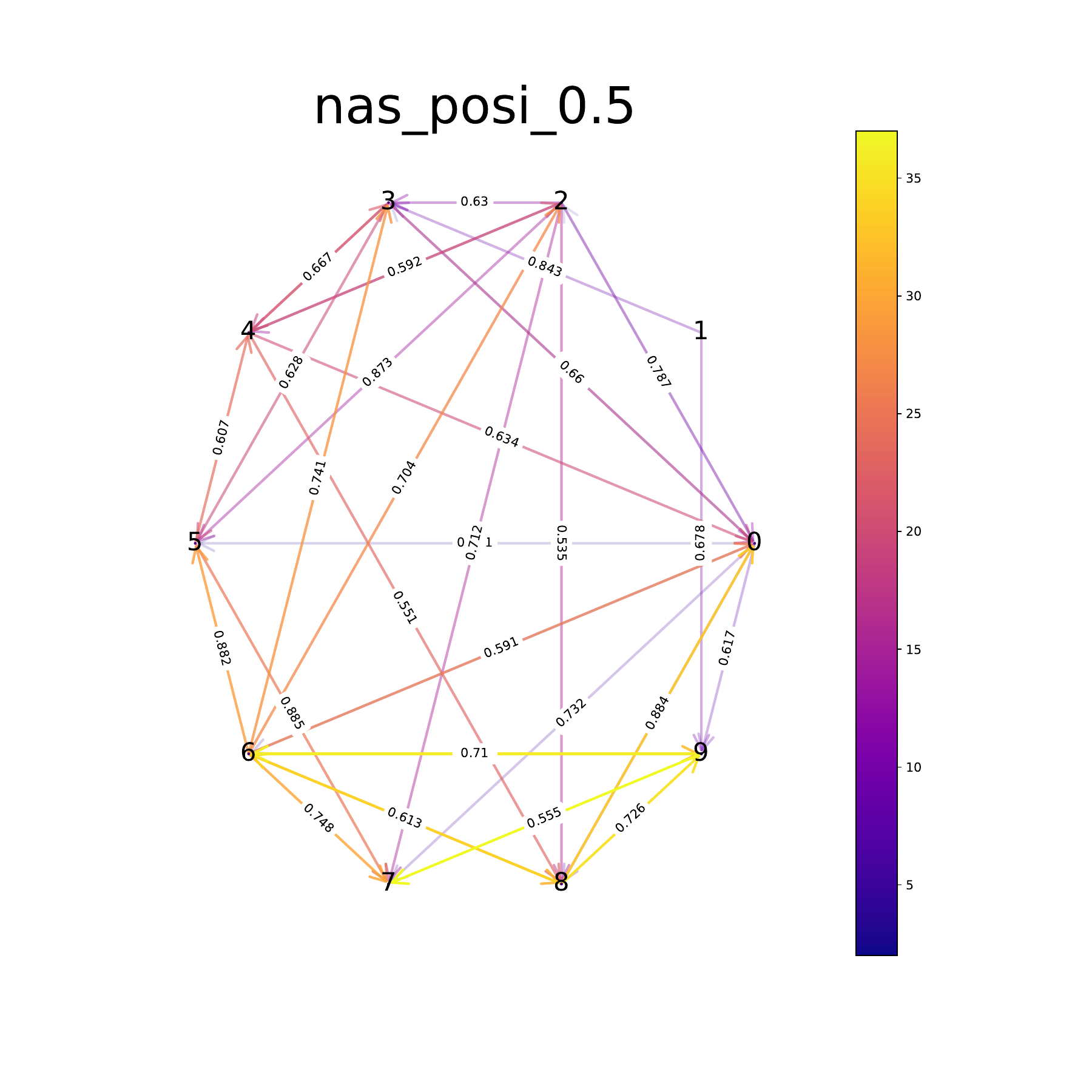}
        \includegraphics[width=0.3\textwidth]{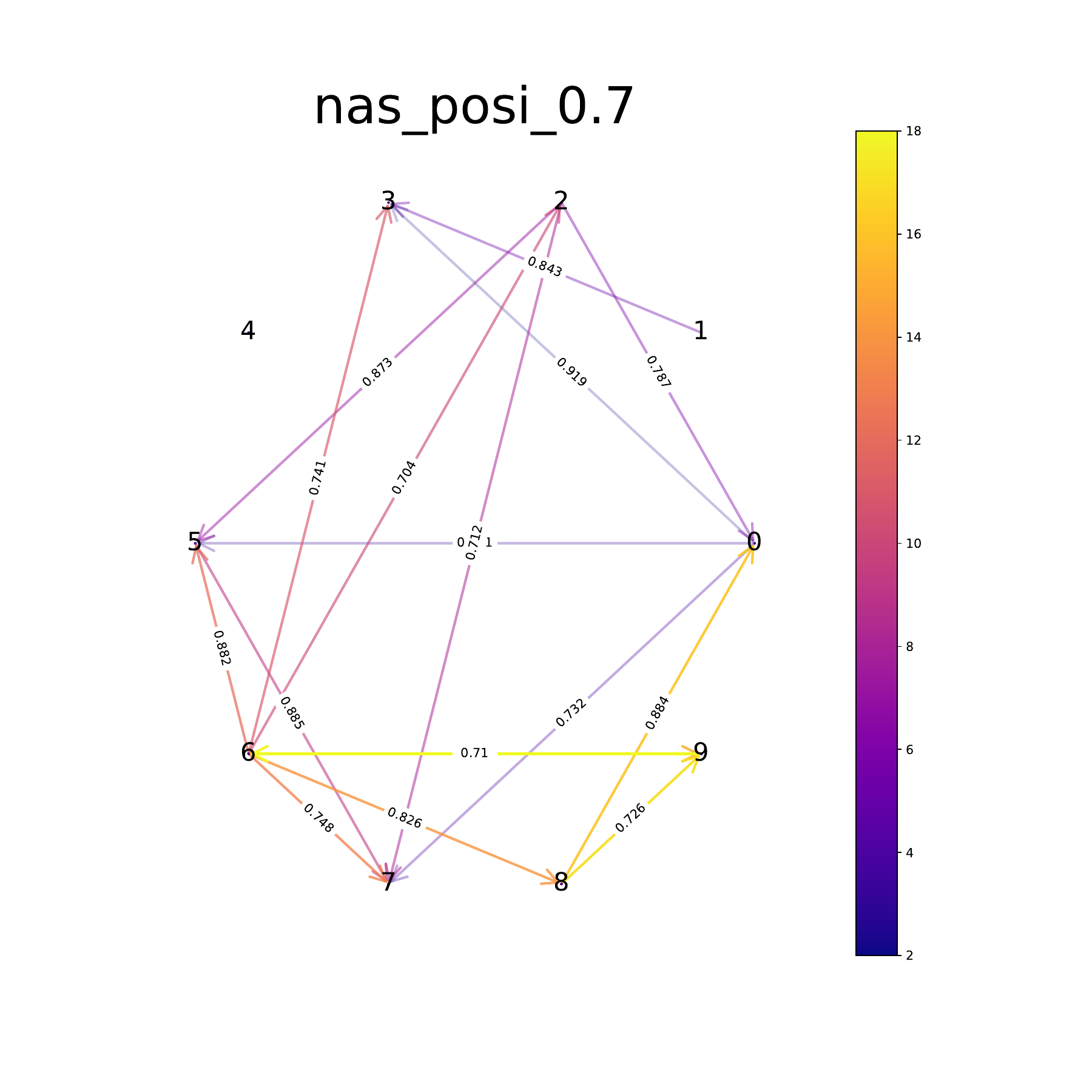}\\
        \includegraphics[width=0.3\textwidth]{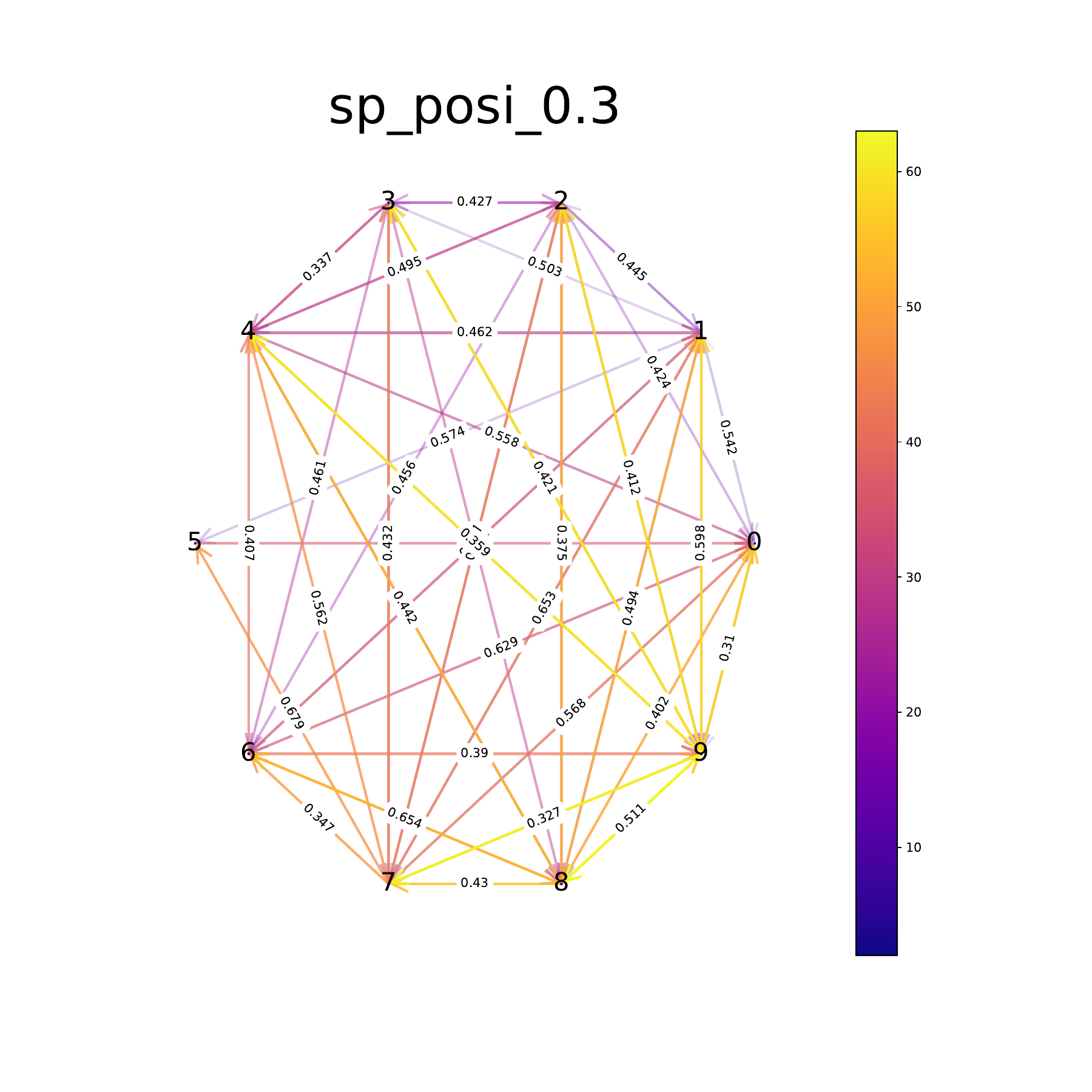}
        \includegraphics[width=0.3\textwidth]{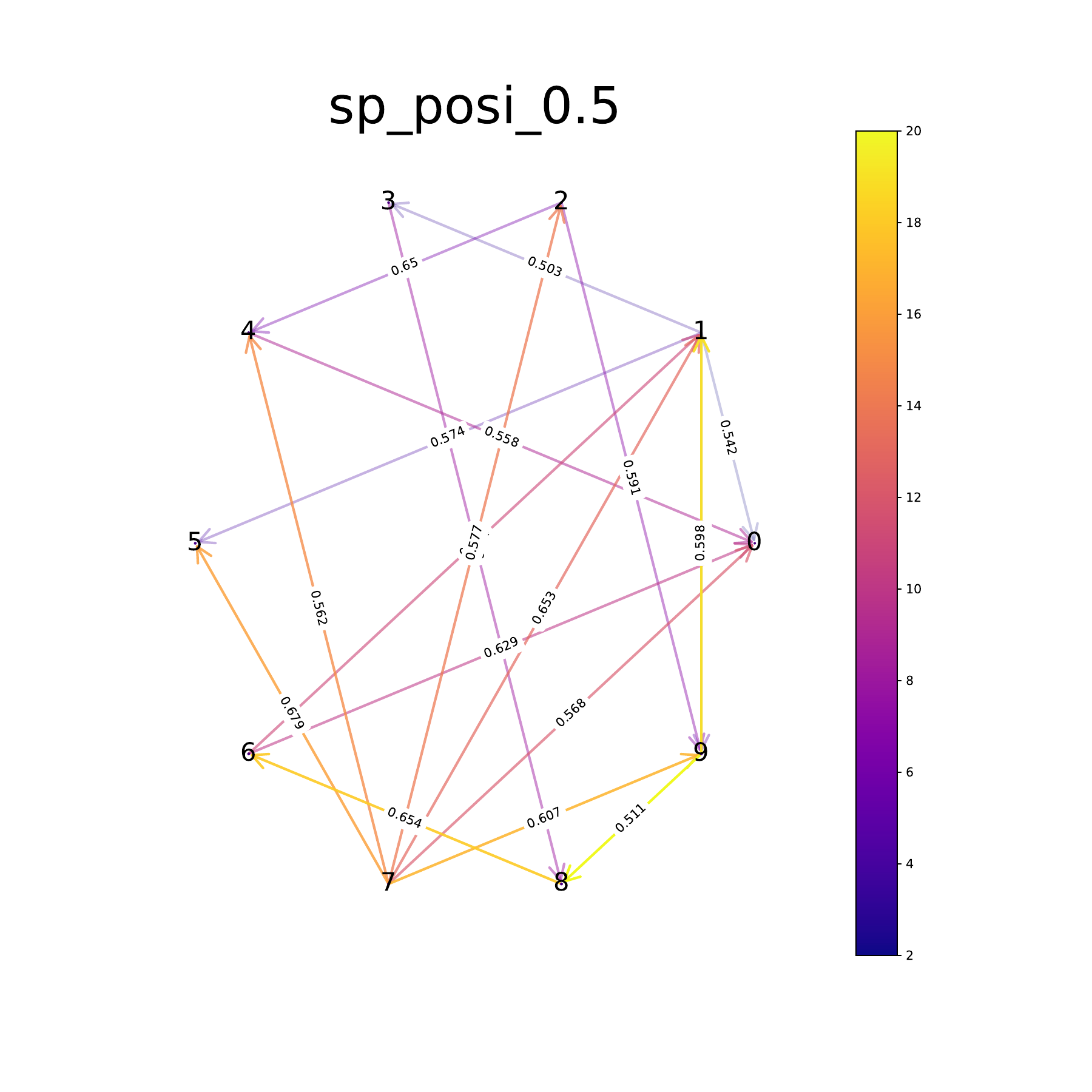}
        \includegraphics[width=0.3\textwidth]{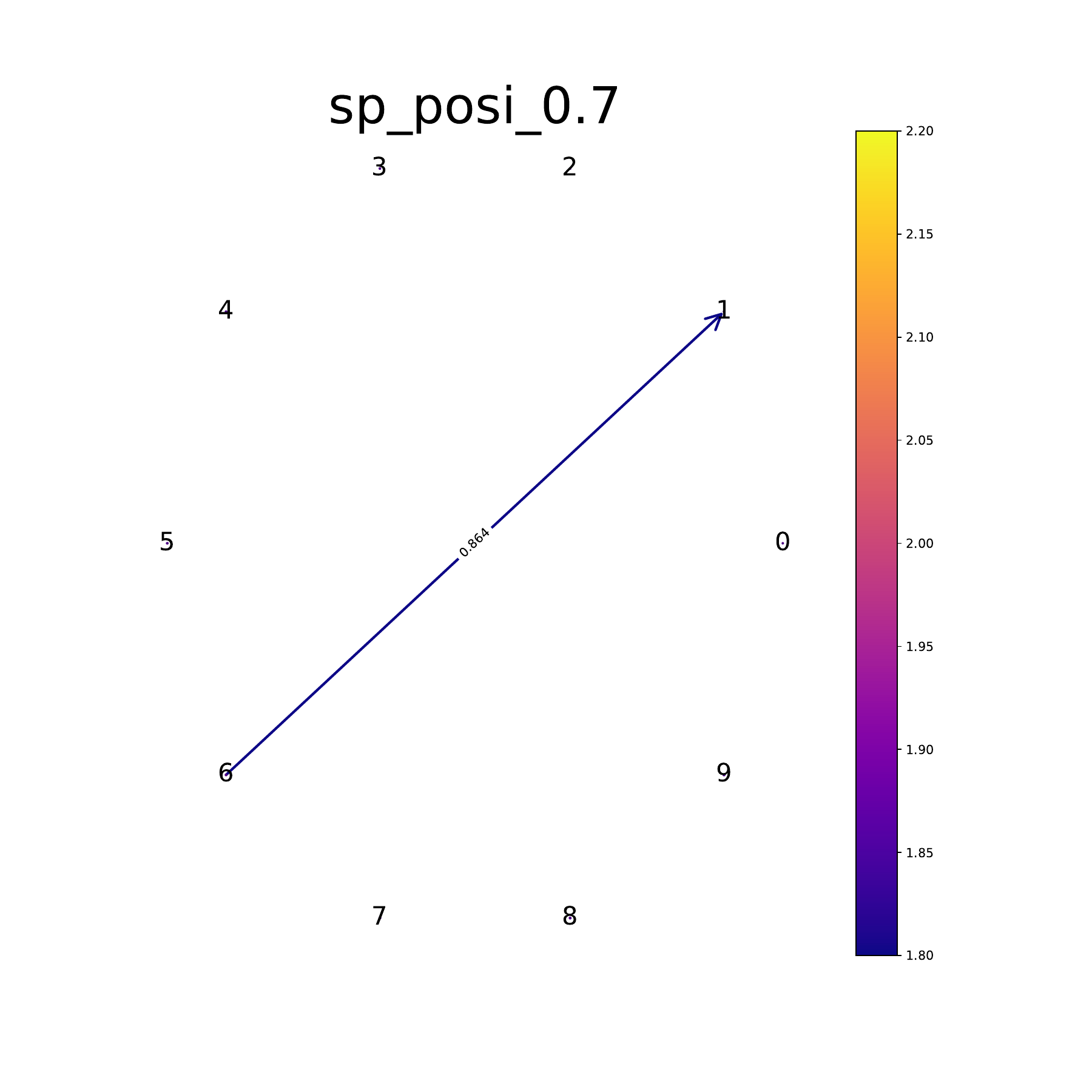}
		\label{Positive relationships in markets}
	\end{minipage}}
    \subfigure[Negative relationships]{
	\begin{minipage}{0.49\linewidth}
        \includegraphics[width=0.3\textwidth]{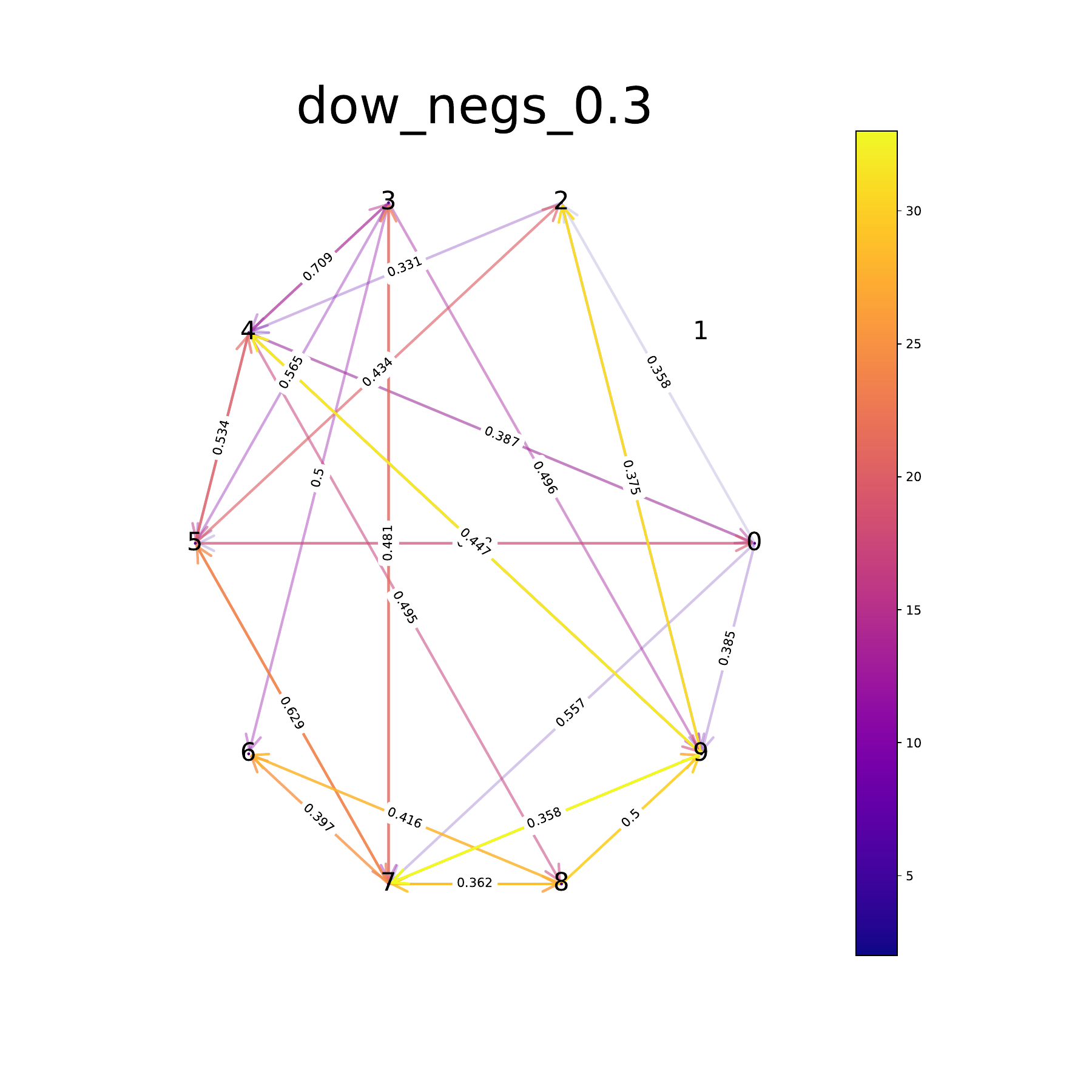}
        \includegraphics[width=0.3\textwidth]{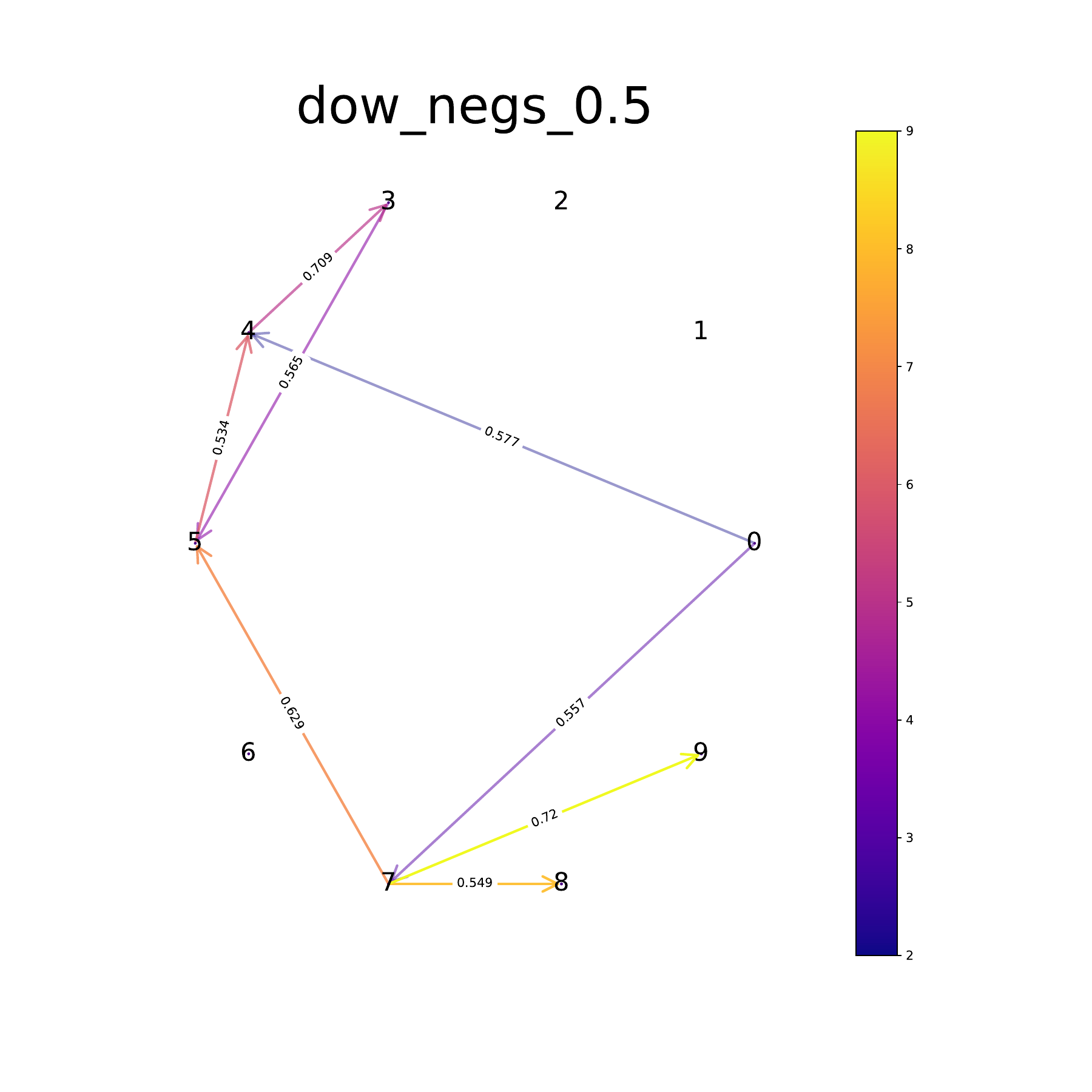} \includegraphics[width=0.3\textwidth]{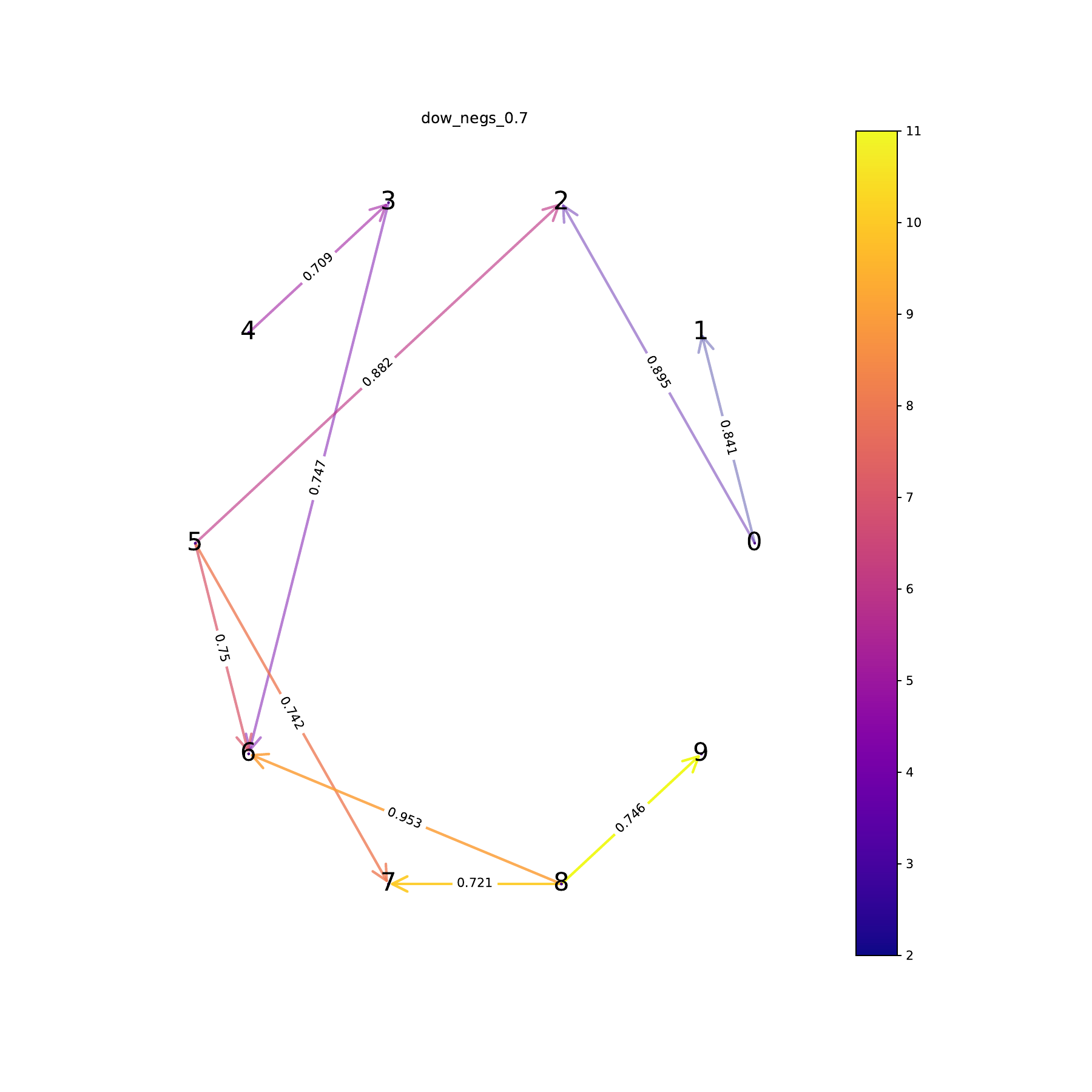}\\
        \includegraphics[width=0.3\textwidth]{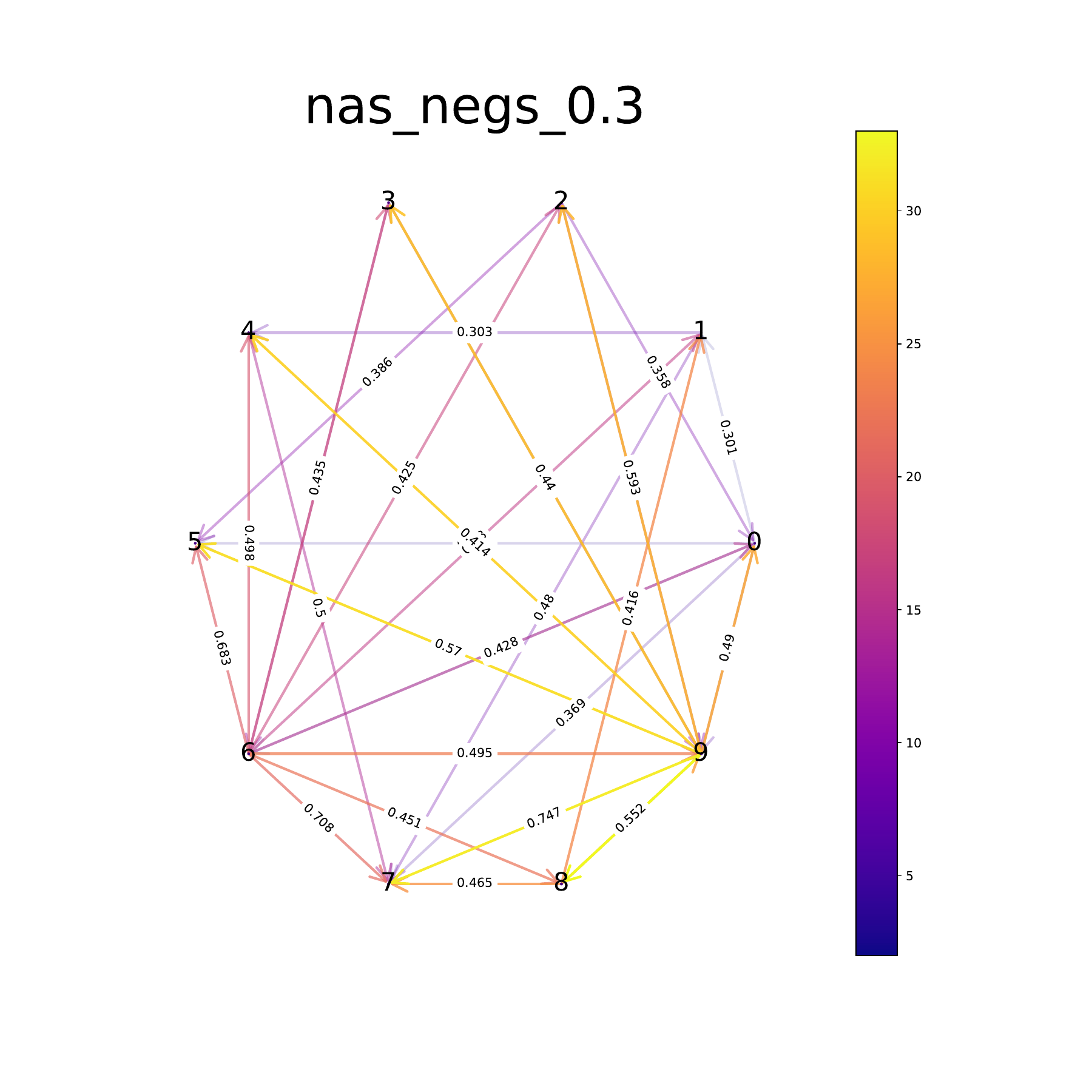}
        \includegraphics[width=0.3\textwidth]{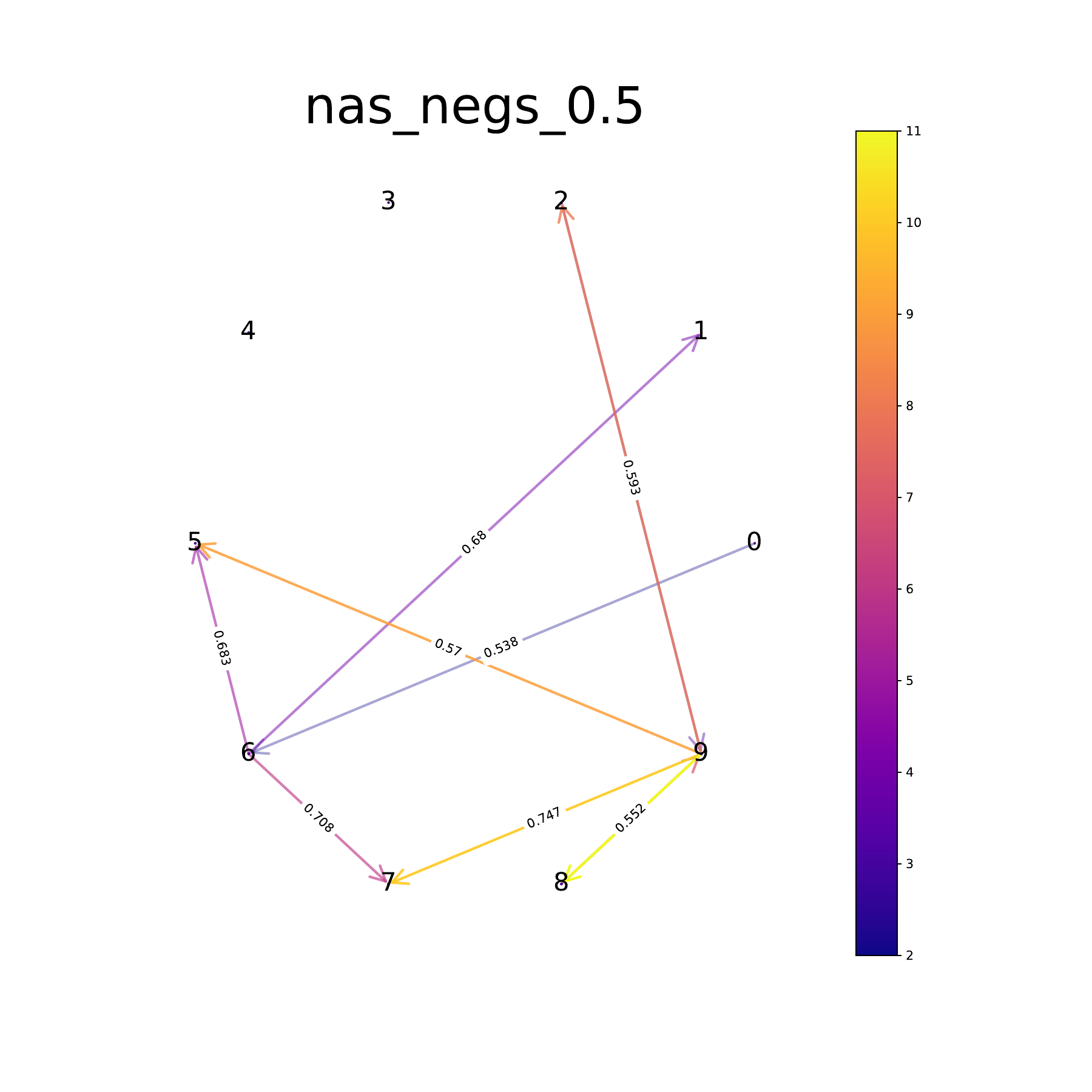}
        \includegraphics[width=0.3\textwidth]{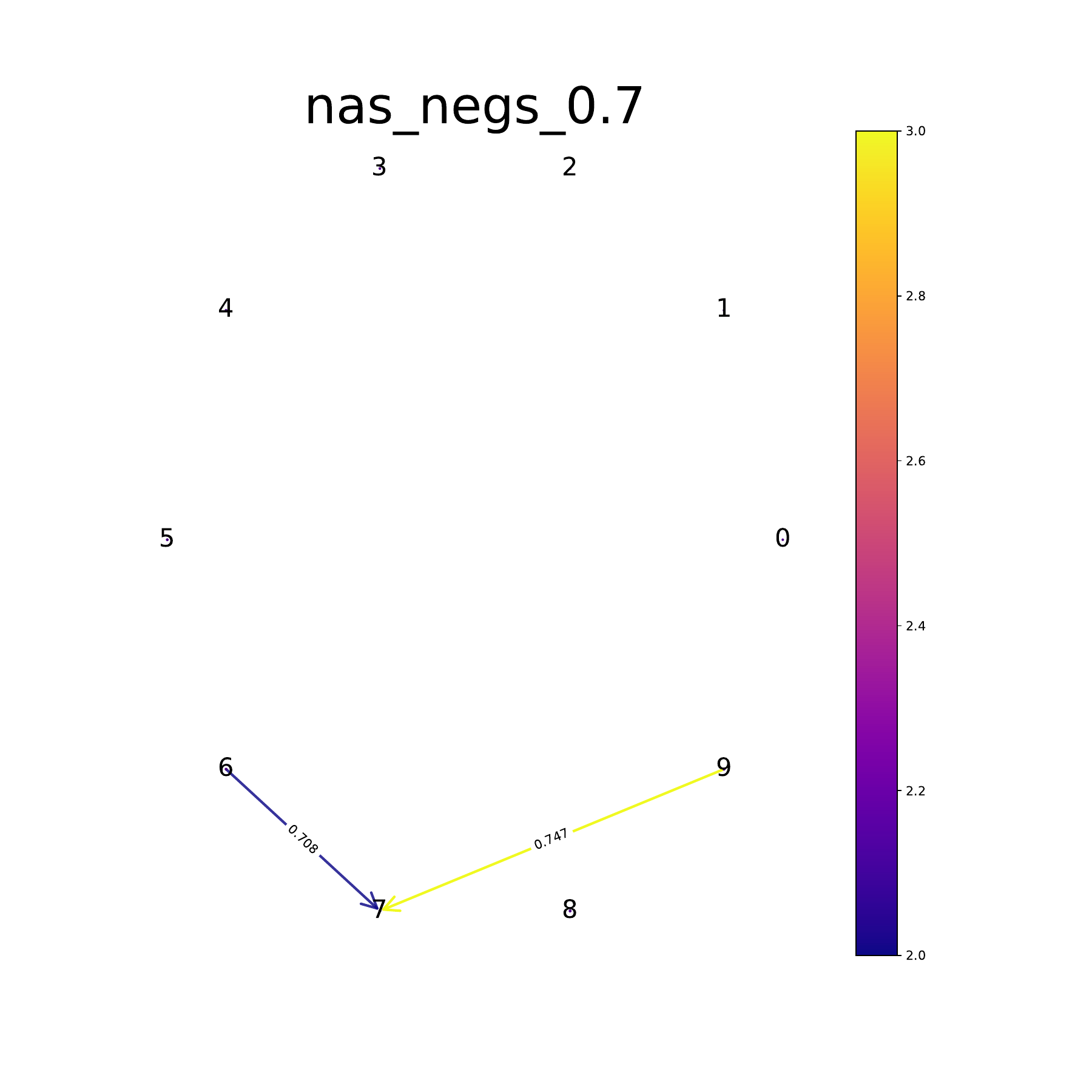}\\
        \includegraphics[width=0.3\textwidth]{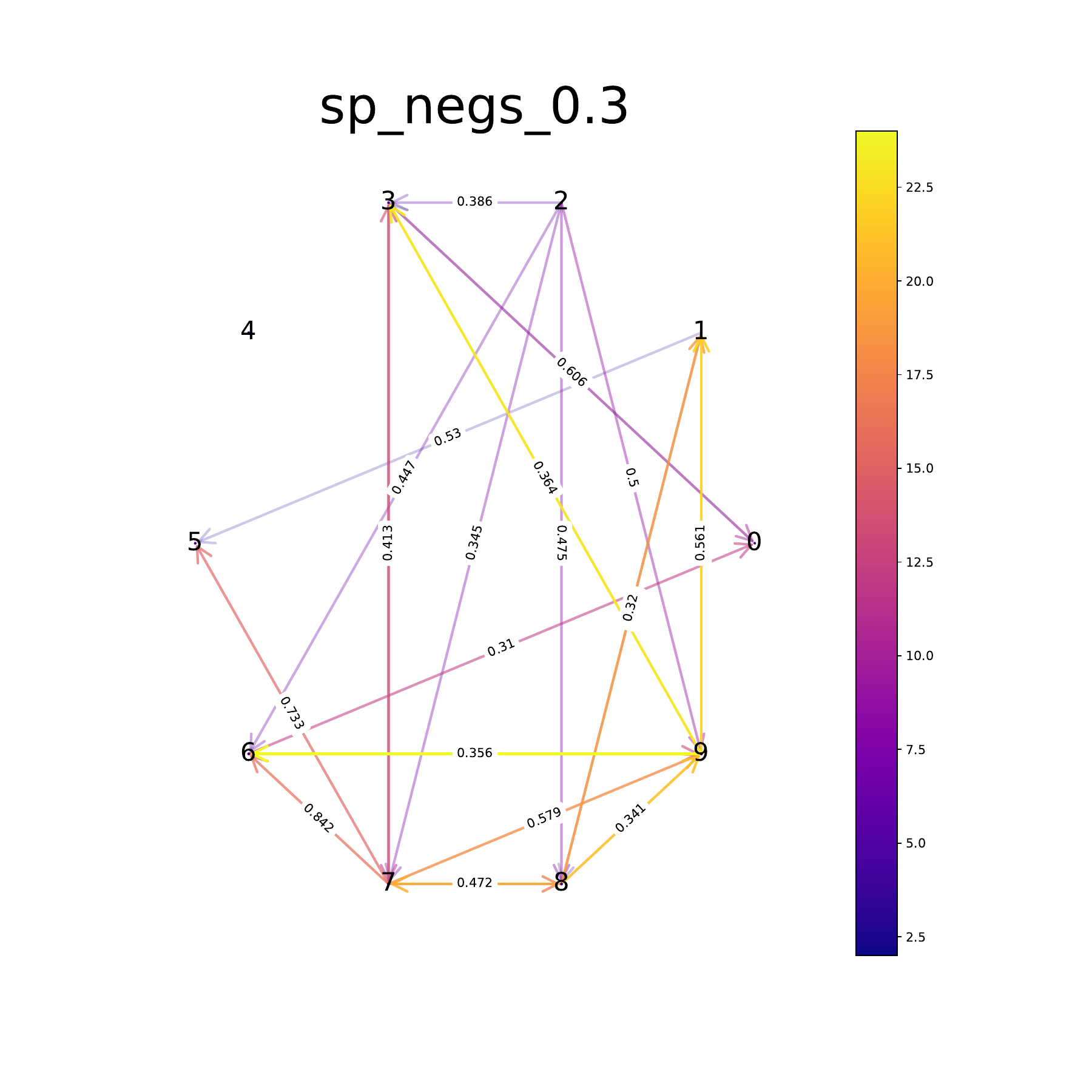}
        \includegraphics[width=0.3\textwidth]{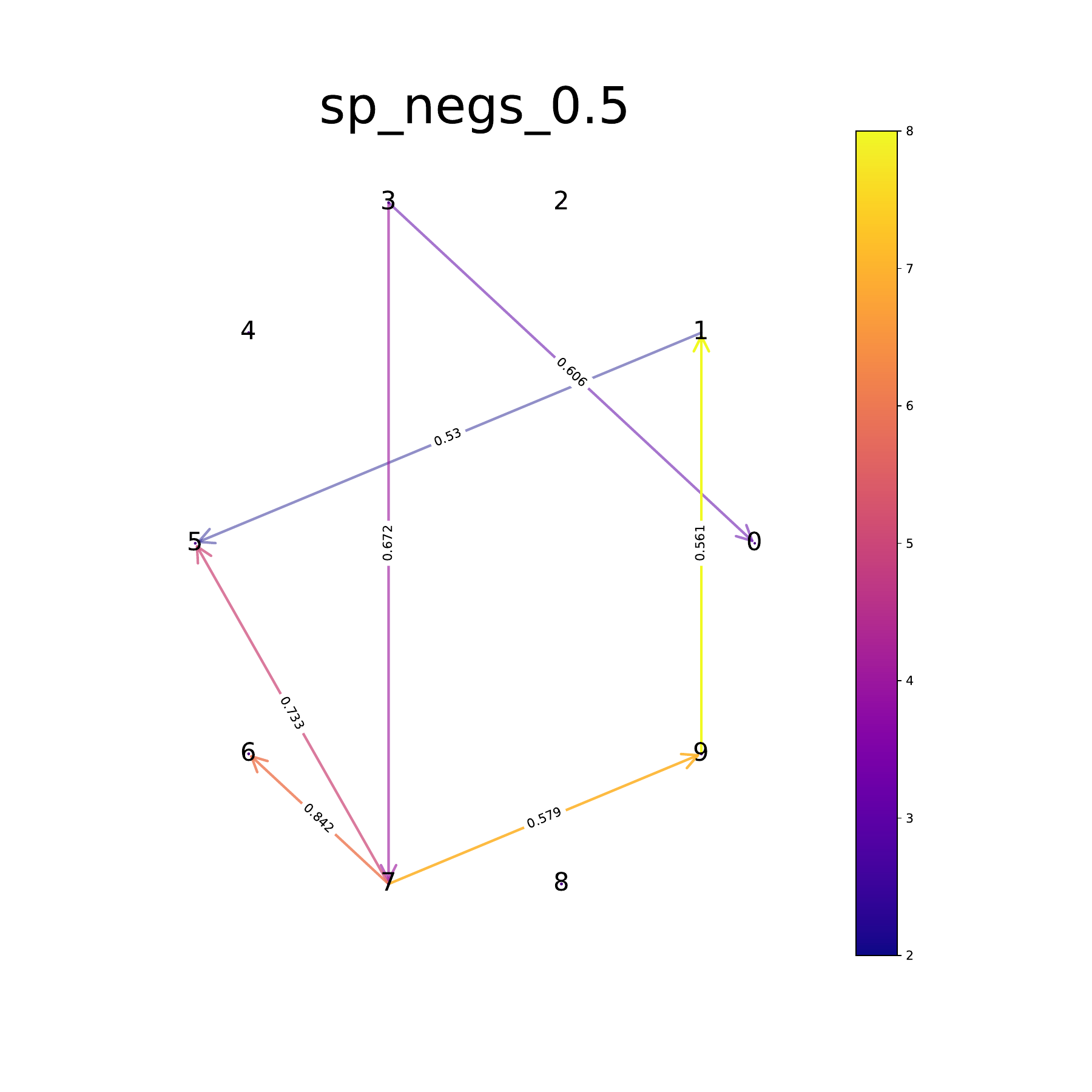}
        \includegraphics[width=0.3\textwidth]{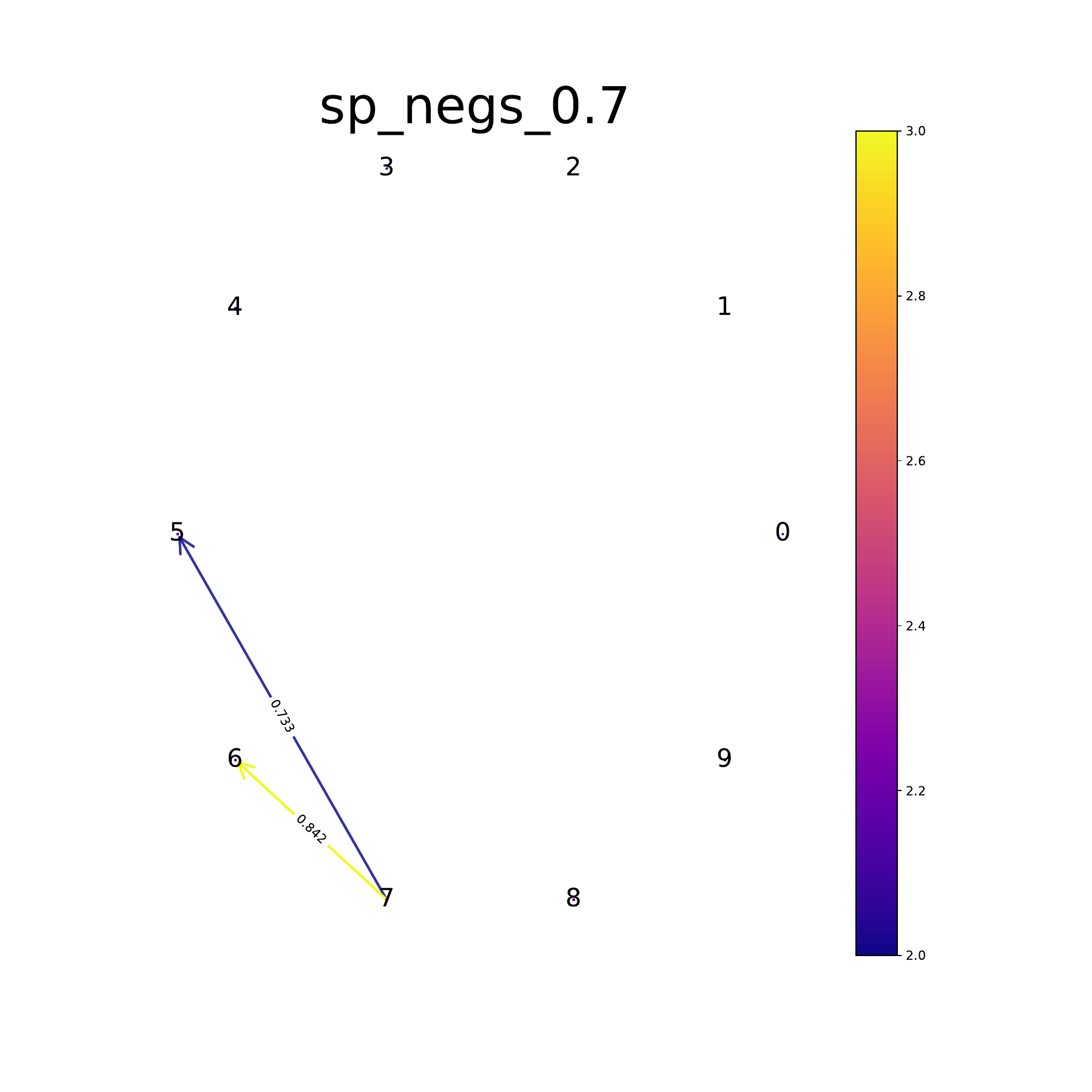}
        \label{Negative relationships in markets}
	\end{minipage}}
 \captionsetup{font=small}
 \caption {From top to bottom, the three columns respectively represent the Dow, NASDAQ and S\&P market, and the three rows from left to right respectively represent the stock correlation scores greater than 0.3, 0.5 and 0.7. Subfigure \ref{Positive relationships in markets} shows the pattern causality analysis of the positive correlation in the three markets, while Subfigure \ref{Negative relationships in markets} is positive correlation.}
\label{Causual_Analysis}
\end{figure}

Figure \ref{Causual_Analysis} displays the positive and negative correlations between stocks using the PC method. In Subfigure \ref{Positive relationships in markets}, the strongest positive correlation is observed in the Dow market, followed by the NASDAQ market and the S\&P market, which has a significantly lower correlation than the first two (shown at the bottom of the panel). This explains why the MBNF model does not significantly outperform the MBPO model in the S\&P market. In the NASDAQ market, although the return of the MBNF model is only slightly higher than the baseline, it has certainly surpassed that of the MBPO model, which does not even reach the baseline.


Figure \ref{Causual_Analysis} \ref{Negative relationships in markets} depicts the negative correlations between stocks, aligning with Sub-figure \ref{Causual_Analysis} \ref{Positive relationships in markets}. The Dow market exhibits the highest negative correlation among the three markets, followed by the S\&P market, and lastly the NASDAQ market. The examination of positive and negative correlations across markets confirms the relevance of the heavy-tailed NF model due to the interdependence and competition between the experimental stocks.

\subsubsection{ Loss Analysis}
We provided numerical analysis of dynamic model loss and the actor neural network loss of SAC-agent in this part to verify the algorithm's efficacy. Note that subsection 4.3.4 contains an analysis of the critic neural network loss of the SAC-agent.


\begin{figure}[htbp]
    \centering
    \subfigure[HNF]{
    \label{NF_model_loss}
    \includegraphics[width=0.49\textwidth]{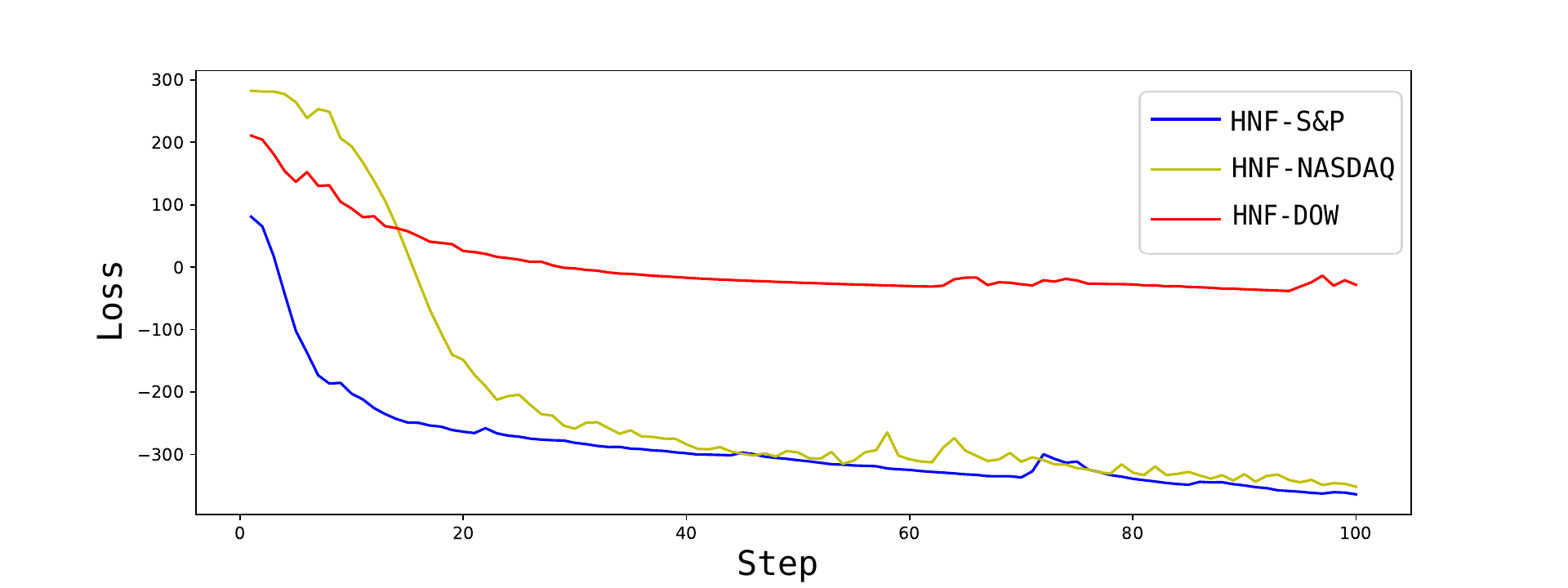}}
    \subfigure[Diagonal Gaussian]{
    \centering
    \label{Gaussians_model_loss}
    \includegraphics[width=0.49\textwidth]{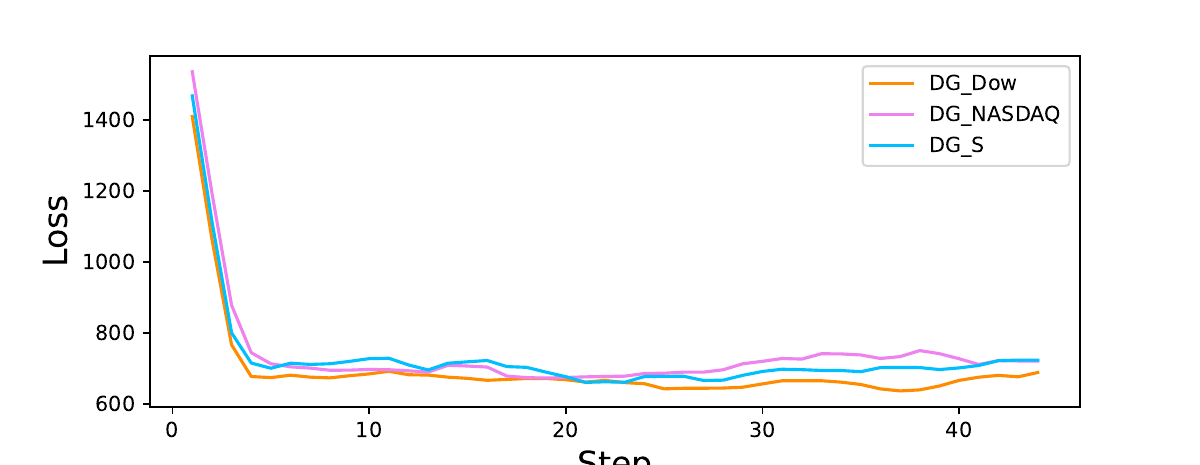}}
\captionsetup{font=small}
\caption{Dynamic Model Loss. (i) and (ii) show the convergence of the model loss in different markets with heavy-tailed NF and diagonal Gaussian as the dynamic model, respectively.}
\label{Dynamic Model Loss}
\end{figure}

We presented separate plots to visualize the performance of the two models as MBNF is optimized by maximizing log-likelihood, while MBPO employs the MSE approach. Figure \ref{Dynamic Model Loss} \ref{NF_model_loss} shows the dynamic model loss with heavy-tailed NF. For all three markets, there is a gradual reduction in loss in the first 20 steps, and by 40 steps, there is consistent convergence. The Dow market shows convergence around 0, while the NASDAQ and S\&P markets converge to a figure around -350. Figure \ref{Dynamic Model Loss}\ref{Gaussians_model_loss}, the diagonal Gaussian dynamic model exhibits rapid loss reduction and converges to about 700 within 20 steps for all three markets. Figure 6 depicts the convergence of the actor network loss for both models in the three markets. After the first 15 steps of training, all solid lines of loss drop smoothly and reach convergence. The convergence values range from 0 to 500, with the diagonal Gaussian model showing significantly higher loss convergence values in the Dow market compared to the other cases.

\begin{figure}[htbp]
    \centering
    \includegraphics[width=0.8\textwidth,height=0.4\textwidth]{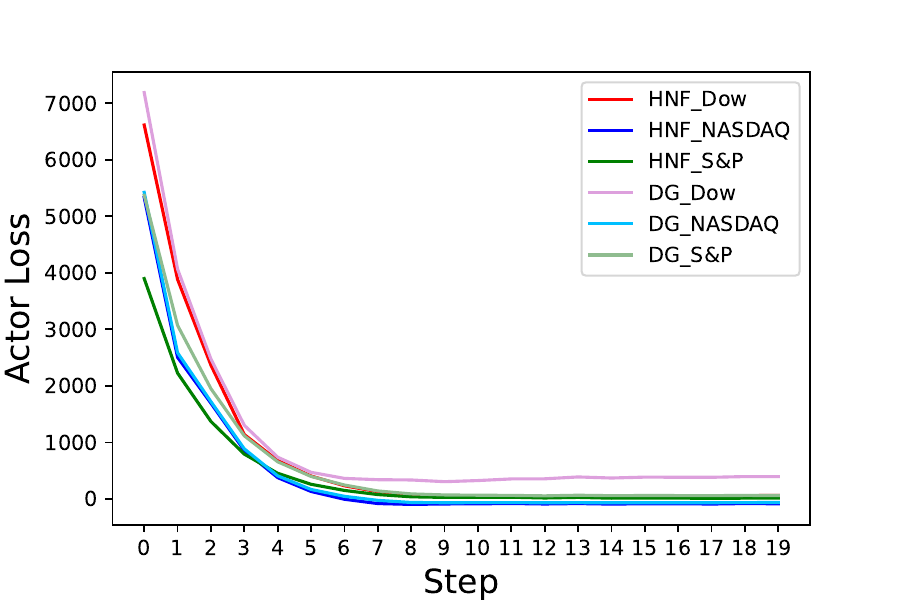}
     \captionsetup{font=small}
    \caption{Actor loss. The X-axis is the training step and Y-axis is the loss value of the actor network of SAC-agent. The HNF in the label denotes heavy-tailed NF model and DG denotes diagonal Gaussian model.}
    \label{actor_loss}
\end{figure}

\subsubsection{ Buffer Analysis}
%
The agent's decision is impacted by the forecasting of the state variable. In order to discover the superior portfolio patterns, we employed the $t$-distributed Stochastic Neighborhood Embedding (t-SNE) technique to analyze the discrepancy in the projected data distribution from the agent training buffer of the MBNF and MBPO algorithms in the S\&P market.


The t-SNE method is a popular nonlinear visualization algorithm that employs a manifold learning method to reduce dimension. It originated in image processing and has since found widespread application in a variety of fields. The basic idea is to map high-dimensional data to a low-dimensional space while preserving the similarity between data points. This is achieved by using a Gaussian probability distribution to measure similarities in the high-dimensional space, and a $t$-distribution probability distribution to measure similarities in the low-dimensional space. The algorithm optimizes the distance between the two probability distributions using the KL divergence, resulting in a sample distribution in the lower dimensional space.

\begin{figure}
    \centering
    \subfigure[Distributions of $S_{t+1}$]{
    \includegraphics[width=0.45\textwidth]{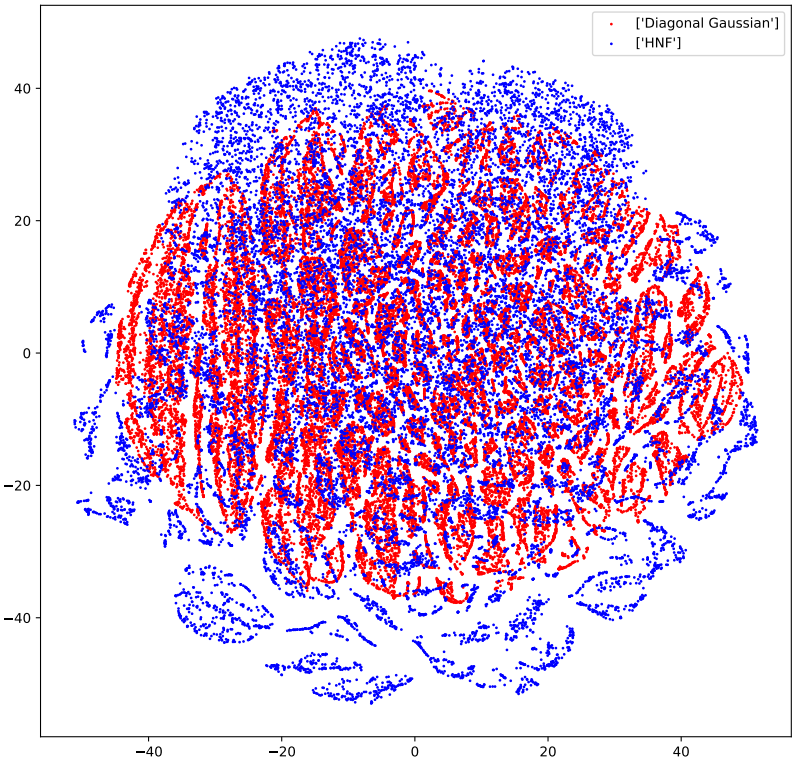}
    \label{predict_$S_{t+1}$}}
    \subfigure[Distributions of $\Delta = S_{t+1} - S_t$]{
    \centering
    \includegraphics[width=0.45\textwidth]{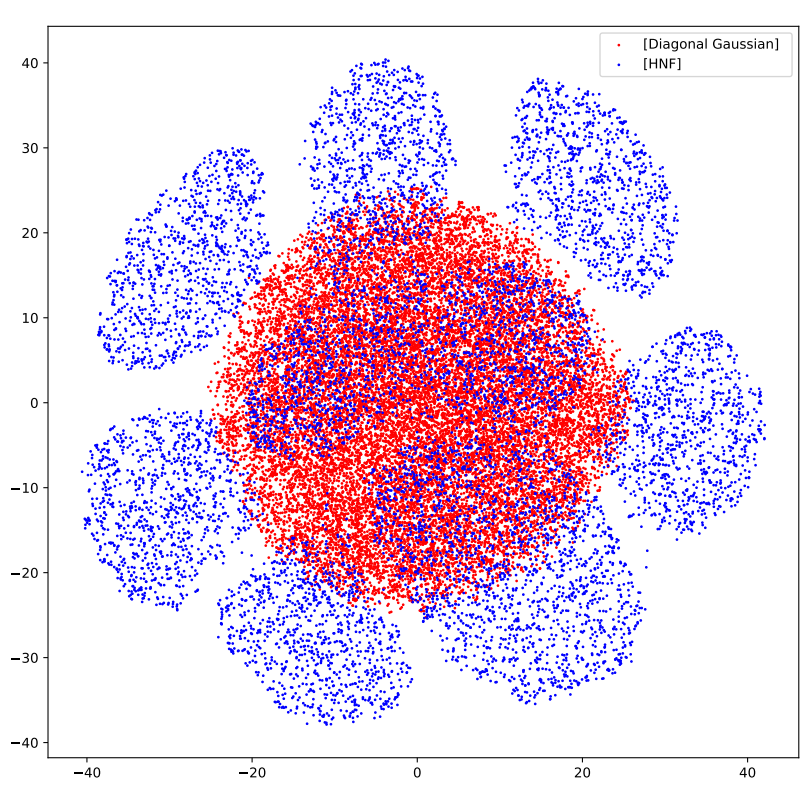}
    \label{predict_delta}}
\captionsetup{font=small}
\caption{The environment buffer and training buffer analysis in S$\&$P Market. Red points represent the reduced dimension graph of state variables simulated by diagonal Gaussian dynamics, and blue points represent the reduced dimension graph of state variables simulated by heavy-tailed NF.}
\label{true buffer and virtual buffer analysis}
\end{figure}

Figure \ref{true buffer and virtual buffer analysis} displays the t-SNE method's environment buffer and training buffer analysis. We can see in 
 Subfigure \ref{predict_$S_{t+1}$} that $s_{t+1}$ generated by heavy-tailed NF has more wild exploration than the diagonal Gaussian dynamic. As seen in Subfigure \ref{predict_delta}, the diagonal Gaussian dynamic model has a strong Gaussian distribution for delta data, whereas the heavy-tailed NF fluxes model has a more mixed Gaussian distribution.

This pattern may suggest that broader sampling and greater variation in distribution of heavy-tailed NF can lead to the production of better trading strategies.


\subsubsection{Eigenvalue Analysis}
As explained in Appendix B, we will illustrate the behavior of the SAC-agent's loss and Hessian eigenvalue to illustrate the proposed model's convergence and validity in two environments, and we have some interesting findings.

We denote the training dataset as $S = \left\{(\mathbf{x}_i, y_i\right)\}_{i=1}^n$, where $ \mathbf{x}_i ,\mathbf{y}_i \in \mathbb{R}^p , n \in \mathbb{Z}^+ \cup \left\{+\infty\right\}$. For all $L$-layer ($ L \geqslant 2$ ) fully-connected neural network $f_{\boldsymbol{\theta}}(x)$, where $\boldsymbol{\theta}$ is the set of all network parameters. The training objective loss function is$ \mathcal{L}_S(\boldsymbol{\theta}) = \mathbb{E}_S \ell(f_{\boldsymbol{\theta}}(\mathbf{x}_i), \mathbf{y}_i)$, where the loss $\ell(f_{\boldsymbol{\theta}}(\mathbf{x}_i), \mathbf{y}_i)$ is differentiable, if the activation function $\sigma(\cdot)$ is differentiable, the Hessian matrix of the objective $ \mathcal{L}_S(\boldsymbol{\theta}) $ is (Note that the following still holds for an activation function like ReLU, as long as unique sub-gradient are assigned for its non-differentiable points):
$$ \nabla_{\boldsymbol{\theta}}^2 \mathcal{L}_S(\boldsymbol{\theta}) = \mathbf{E}_s \nabla_{\boldsymbol{\theta}}^2 \ell(f_{\boldsymbol{\theta}}(\mathbf{x}_i,\mathbf{y}_i) , \quad \quad \quad \mathbf{H} \triangleq \nabla_{\boldsymbol{\theta}}^2 \mathcal{L}_S(\boldsymbol{\theta}) $$
Here, loss $\ell(f_{\boldsymbol{\theta}}(\mathbf{x}_i), \mathbf{y}_i)$ is Mean Squared Error $\ell(\mathbf{z}, \mathbf{w})=\|\mathbf{z}-\mathbf{w}\|_2^2$ for every $\mathbf{z}, \mathbf{w} \in \mathbb{R}^p$; $\| \cdot\|_2$ is Euclidean norm, and ReLU employed for activation layer in both critic and actor neural network of SAC-agent.

For the eigenvalues, sharpness, and the remaining set of eigenvalues of matrix $\mathbf{H}$, respectively, we use $\Lambda$, $\lambda_s$, and $\Lambda_r$. The following definitions are more detailed.
$$
 \Lambda \triangleq \{\text{Eigenvalues of matrix} \mathbf{H} \},
\lambda_s \triangleq \{\text{The maximum value of the set} \Lambda \},
\Lambda_r \triangleq \Lambda \setminus \{ \lambda_s \} \\
$$



\begin{figure}[htbp]
    \centering
    \subfigure[With technical indicators]{
    \label{dow_hessian_egi_True}
    \includegraphics[width=6cm]{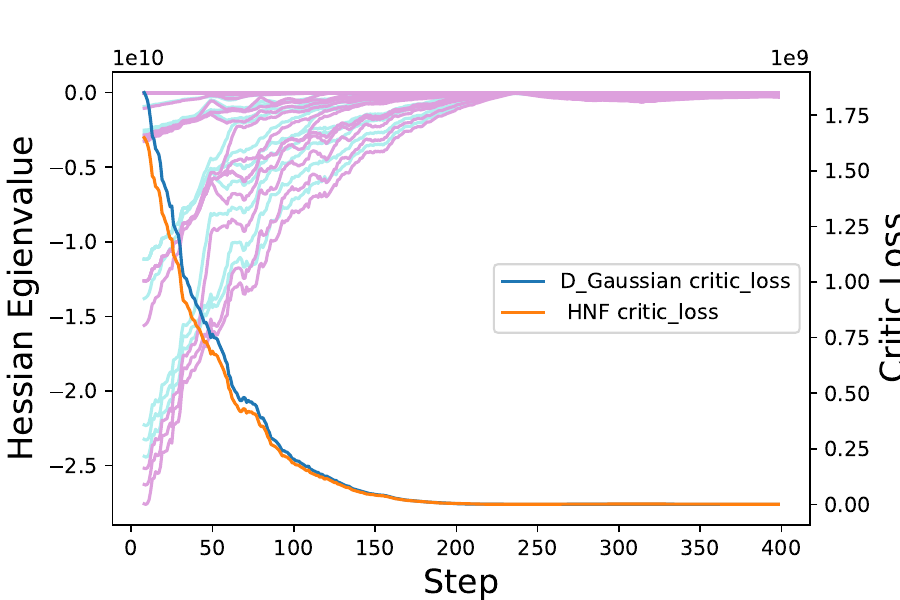}}
    \subfigure[Without technical indicators]{
    \quad
    \centering
    \label{dow_hessian_egi_False}
    \includegraphics[width=6cm]{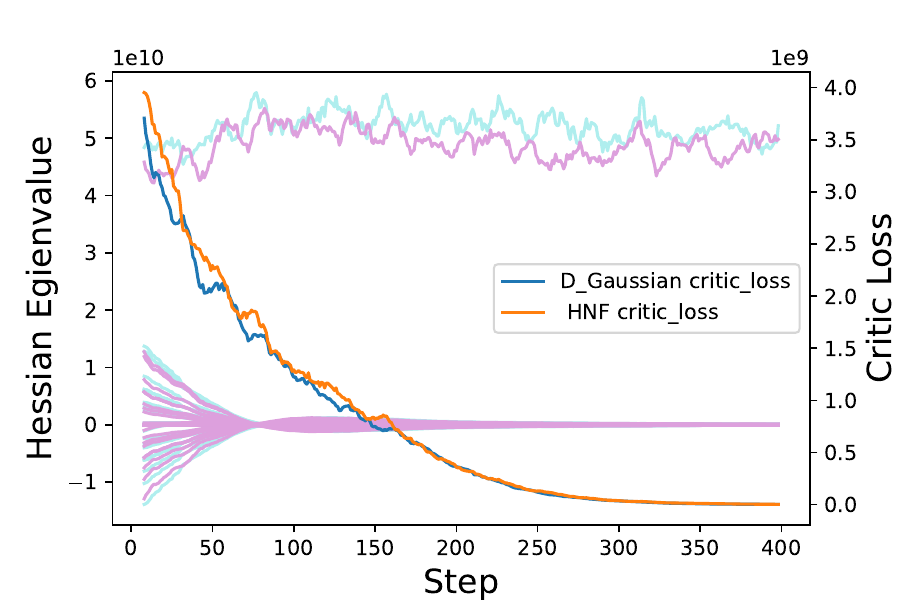}}
\captionsetup{font=small}
\caption{The behavior study of the critic network's eigenvalues and its loss in two environments. Subfigure (i) corresponds to the case that technical indicators are included in the state of environment, whereas (ii) depicts an absent case. The left Y-axis represents the eigenvalues of the matrix $\mathbf{H}$ of the critic network, the right Y-axis is the network loss value, and the X-axis represents the number of training steps. The plum solid lines and the orange solid line represent the eigenvalues and loss values for the heavy-tailed NF, respectively, while the pale turquoise solid lines and the blue solid line are obtained from the diagonal Gaussian model. Note that the eigenvalues of $\mathbf{H}$ are calculated by the NumPy library.}
\label{hessian_egienvalue}
\end{figure}

Figure \ref{hessian_egienvalue} illustrates the relationship between the convergence of the SAC-agent's critic network and $\Lambda$ in the Dow market when the diagonal Gaussian model and the heavy-tailed NF model are used as dynamic models in two different situations – with and without the use of technical considerations.

Figure \ref{hessian_egienvalue} \ref{dow_hessian_egi_True} corresponds to the state containing technical indicators. As we cannot distinguish between $\lambda_s$ and $\Lambda_r$, we established an association between the convergence behavior of loss and the collective behavior of the set $\Lambda$. It may be broken down into three stages specifically. Stage 1: In the first 100 steps of training, the loss value falls significantly and quickly and $\Lambda$ converges to the X-axis. Stage 2: Loss slowly converges, while $\Lambda$ value keeps ascending. Stage 3: After 250 training steps, both loss and eigenvalues reach convergence and set $\Lambda$ converges to a band around the X-axis. Notice that the loss reaches convergence before the eigenvalues do. Throughout the convergence process, the negative eigenvalues keep converging to near 0, which suggests that there may be one attractor in the network, and the final broadband convergence indicates that our model finally reaches near the attractor. 

Figure \ref{hessian_egienvalue}\ref{dow_hessian_egi_False} shows the case where no technical indicators are used, with a totally different situation. As observed, there is a sizable difference between $\lambda_s$ and the $\Lambda_r$, and the two act quite differently throughout training. We established a substantial association between the convergence behavior of loss and the collective behavior of the set $\Lambda_r$, illustrating the fact that our data shows no correlation between loss values and $\lambda_s$. It may be broken down into four stages specifically. Stage 1: In the first 80 steps of training, the loss value falls significantly and quickly as $\Lambda_r$ as a whole converges to X-axis. Stage 2: Following the first aggregation, the set $\Lambda_r$ somewhat diverges and slowly converges, while the loss value keeps dropping. Stage 3: After 250 training steps, set $\Lambda_r$ converges for a second time and stabilizes in a band around the X-axis, while the loss value gradually lowers and converges a little later than the eigenvalue. Stage 4: Set $\Lambda_r$ and the loss both achieve the convergence state after 300 training steps. The neural network has many saddle points, as shown by the early training's numerous positive and negative eigenvalues, and by the time the eigenvalues converge to around 0, the model parameters have reached a smooth zone. The absolute values of the loss and eigenvalue obtained by the NF model are bigger than those of the G model, which is different from the first case. Despite the fact that we are unsure of the physical reasons for the huge difference between \ref{dow_hessian_egi_True} and \ref{dow_hessian_egi_False}, we do know that two cases have reliable results.

\section{Conclusion}
In this paper, we utilized a heavy-tailed preserving normalizing flow to simulate the high-dimensional joint probability of the complex trading environment and developed a model-based reinforcement learning framework called MBNF to gain a better understanding of the intrinsic mechanisms of quantitative trading. We give the portfolio back-testing experiments with various stocks from three different financial markets (Dow, NASDAQ and S$\&$P 500). Our results demonstrate that Dow outperforms the other two markets according to various evaluation metrics, even during the unpredictable crisis of the COVID-19 pandemic. Furthermore, we explored the explanation of our RL algorithm through various methods. Firstly, we utilized the pattern causality method to study the interactive relation among different stocks of the environment. Among the three markets, Dow market has the highest correlation, followed by S$\&$P market and NASDAQ market. This also explains why, in S$\&$P market, the proposed model does not have a significant improvement over the MBPO model. Secondly, we analyzed the dynamic loss and actor loss to ensure the convergence of the two strategies. By visualizing high dimensional state transition data from training and environment buffer with t-SNE, we uncovered some effective patterns of better portfolio optimization strategies. Finally, we also utilized eigenvalue analysis in the two environments to study the convergence properties of the environment neural network model which gave some interesting findings.

The results illustrate that heavy-tailed normalizing flows will capture the correlation between high-dimensional stock data, facilitating the decision-making process of many automated trading systems.

However, there are still some challenges that are open to us. For example, the environment from real financial markets is complex and hard to fully simulate, considering the non-stationarity of the stock data, and unpredictable hidden causal factors. Additionally, since there is a close connection between RL and stochastic optimal control, it is unclear how to demonstrate the uniqueness and properties of the solution in an appropriate functional space once the problem is set up as a stochastic optimal control problem. Moreover, the analysis of the generalised error bound of model-based RL is also essential for us to design more adaptive and stable RL algorithms.

\section*{ACKNOWLEDGMENTS}
This work was supported by NSFC grants 11771449 and The Science and Technology Innovation 2030—Brain Science and
Brain-Inspired Intelligence Project (2021ZD0201301).


\appendix
\section{Data Exploration with Stable Distribution} 

Here we put the parameters estimation of $S_\alpha(\sigma, \beta, \mu)$ distribution for the Dow and S$\&$P market.

\begin{minipage}{\textwidth}
    \begin{minipage}[t]{0.45\textwidth}
    \makeatletter\def\@captype{table}
    \setlength\tabcolsep{2.8pt}
    \centering
    \begin{tabular}{ccccc} 
    \hline
    \multicolumn{5}{c}{Dow}                           \\ 
    \hline
    ID & $\alpha$  & $\beta$    & $\mu$    & $\sigma$   \\
    \hline
    HD         & 1.5788  & 0.0102  & 0.0697 & 0.4735  \\
    JNJ        & 1.6288  & 0.0035  & 0.0472 & 0.3601  \\
    KO         & 1.7205  & -0.1988 & 0.0231 & 0.1702  \\
    MCD        & 1.7125  & -0.1793 & 0.0729 & 0.4343  \\
    MSFT       & 1.5845  & 0.1665  & 0.0064 & 0.2581  \\
    NKE        & 1.5628 & 0.0338  & 0.0162 & 0.2681  \\
    PG         & 1.7060   & 0.0491  & 0.0150  & 0.3149  \\
    UNH        & 1.5713  & 0.0602  & 0.0584 & 0.5832  \\
    V          & 1.6152  & -0.0245 & 0.0493 & 0.3656  \\
    WMT        & 1.6951  & -0.1042 & 0.0373 & 0.3314  \\
    \hline
    \end{tabular}
    \caption{Dow Market}
    \label{DOW_fitting_appendix}
    \end{minipage}
    \begin{minipage}[t]{0.45\textwidth}
    \makeatletter\def\@captype{table}
    \setlength\tabcolsep{2.8pt}
    \centering
    \begin{tabular}{ccccc} 
    \hline
    \multicolumn{5}{c}{S\&P}                          \\
        \hline
    ID & $\alpha$  & $\beta$    & $\mu$    & $\sigma$   \\
    \hline
    ABT        & 1.6228 & -0.0252 & 0.0266 & 0.2092  \\
    ACN        & 1.6421 & 0.0701  & 0.0758 & 0.5101  \\
    GOOGL      & 1.5915 & -0.0262 & 0.3808 & 3.7552  \\
    ISRG       & 1.6633 & 0.0550   & 0.1117 & 1.4620   \\
    ITW        & 1.7085 & -0.0786 & 0.0671 & 0.4572  \\
    JNJ        & 1.6288 & 0.0035  & 0.0472 & 0.3691  \\
    PM         & 1.7263 & -0.1720  & 0.0565 & 0.3753  \\
    ROL        & 1.6333 & 0.0487  & 0.0059 & 0.0604  \\
    SYK        & 1.7110  & -0.0975 & 0.0841 & 0.4963  \\
    WMT        & 1.6951 & -0.1042 & 0.0373 & 0.3314  \\
    \hline
    \end{tabular}
    \caption{S\&P Market}
    \label{SP_fitting_appendix}
    \end{minipage}
\end{minipage}

\section{Egienvalue Analysis}
The eigenvalues of Hessian matrix arising in ML models have close relations with model loss. For example, Yaoyu Zhang et al. (2021)~\cite{zhang2021embedding} defined the critical points using the eigenvalues of the Hessian matrix, then utilized the critical points to examine the loss landscape of a deep neural network and developed the embedding principle. There are many theoretical perspectives [~\cite{chaudhari2019entropy}, ~\cite{jacot2019asymptotic}] and empirical perspectives [~\cite{dong2019hawq}, ~\cite{yao2021adahessian}] on the eigenspectra and local loss curvature of Hessian matrices. The largest eigenvalue of the Hessian matrix of the objective is known as “sharpness” in the literature. Zhouzi Li et al. (2022)~\cite{li2022analyzing} divided the dynamics of loss into four phases based on the change in the sharpness value to examine the Edge of Stability (EOS) phenomenon from a theoretical and empirical standpoint. Zhenyu Liao et al. (2021) ~\cite{liao2021hessian} demonstrated that, depending on the data attribute, the model, and the loss function, the Hessian matrix can have fundamentally distinct spectral behaviors by utilizing deterministic equivalent methodologies to produce a precise depiction of the Hessian eigenspectra for realistic nonlinear models. Sagun (2016) et al. ~\cite{sagun2016eigenvalues} found that the Hessian matrix's eigenvalues in deep learning are tightly connected to the data, the model, the training procedures, and other variables. The eigenvalues also, to a certain degree, represent the complexity of the data.

\bibliographystyle{unsrt}
\bibliography{main}

\end{document}